\documentclass[prd,aps,twocolumn,a4paper,showkeys,nofootinbib,floatfix]{revtex4-1}

\usepackage{graphicx,psfrag}
\usepackage{mathrsfs}
\usepackage{amsmath,amsfonts,amssymb}
\usepackage{multirow}
\usepackage{comment}
\usepackage{ulem}
\usepackage{hyperref}
\usepackage{enumitem}
\usepackage{morefloats}

\newcommand{\be}{\begin{equation}}
\newcommand{\ee}{\end{equation}}
\newcommand{\bea}{\begin{eqnarray}}
\newcommand{\eea}{\end{eqnarray}}
\newcommand{\bel}{\begin{align}}
\newcommand{\eel}{\end{align}}

\def\dee{{\rm d}}

\def\Msun{{\rm M_{\odot}}}
\def\GMc2{{\rm G M_{\odot} c^{-2}}}

\def\M{\mathcal{M}}

\def\Mo{{\rm M_{\odot}}}
\def\Mmax{M_\text{TOV}^\text{max}}
\def\Mthr{M_\text{thr}}
\def\kthr{k_\text{thr}}
\def\Lam{\Lambda}
\def\TEOB{\tt TEOBResumS}
\def\TF2{\tt TaylorF2}
\def\IMRP{\tt IMRPhenomPv2NRtidal}
\def\IMRD{\tt IMRPhenomDNRtidal}
\def\SEOB{\tt SEOBNRT}
\def\PPC{P_{\rm PC}}

\usepackage{pifont} 
\usepackage{color}
\definecolor{cyan}{rgb}{0,0.9,0.9}
\definecolor{orange}{rgb}{0.9,0.5,0}
\definecolor{magenta}{rgb}{1,0,1}
\definecolor{purple}{rgb}{0.8,0.4,0.8}
\definecolor{gray}{rgb}{0.8242,0.8242,0.8242}

\begin{document}
\title{Inferring prompt black-hole formation in neutron star mergers from gravitational-wave data}

\author{Michalis \surname{Agathos}$^{1}$}
\author{Francesco \surname{Zappa}$^{1}$}
\author{Sebastiano \surname{Bernuzzi}$^{1}$}
\author{Albino \surname{Perego}$^{2,3}$}
\author{Matteo \surname{Breschi}$^{1}$}
\author{David \surname{Radice}$^{4,5}$}

\affiliation{${}^1$Theoretisch-Physikalisches Institut, Friedrich-Schiller-Universit{\"a}t Jena, 07743, Jena, Germany}
\affiliation{${}^2$Dipartimento di Fisica, Universit\'a di Trento, Via Sommarive 14, 38123 Trento, Italy}
\affiliation{${}^3$Istituto Nazionale di Fisica Nucleare, Sezione di Milano-Bicocca, Piazza della Scienza 20100, Milano, Italy}
\affiliation{${}^4$Institute for Advanced Study, 1 Einstein Drive, Princeton, NJ 08540, USA}
\affiliation{${}^5$Department of Astrophysical Sciences, Princeton University, 4 Ivy Lane, Princeton, NJ 08544, USA}

\date{\today}

\begin{abstract}
  The gravitational-wave GW170817 is associated to the inspiral
  phase of a binary neutron star coalescence event.
  The LIGO-Virgo detectors sensitivity at high frequencies was not sufficient to detect the
  signal corresponding to the merger and post-merger phases.
  Hence, the question whether the merger outcome was a prompt black hole
  formation or not must be answered using either the pre-merger
  gravitational wave signal or electromagnetic counterparts.
  In this work we present two methods to infer the probability of prompt black hole
  formation, using the analysis of the inspiral gravitational-wave signal.
  Both methods combine the posterior distribution from the gravitational-wave data analysis
  with numerical relativity results. One method relies on the use of
  phenomenological models for the equation of state and on the estimate of
  the collapse threshold mass. The other is based on the estimate of
  the tidal polarizability parameter $\tilde{\Lambda}$ that is correlated in an 
  equation-of-state agnostic way with the prompt BH formation.
  We analyze GW170817 data and find that the two methods 
  consistently predict a probability of $\sim 50$-$70\%$ for
  prompt black hole formation, which however may significantly decrease below $10\%$
  if the maximum mass constraint from PSR J0348+0432 or PSR J0740+6620 is imposed.
  %
\end{abstract}


\maketitle




\section{Introduction}
\label{sec:intro}


The gravitational-wave (GW) signal GW170817, detected by the LIGO-Virgo 
detector network~\cite{TheLIGOScientific:2014jea,TheVirgo:2014hva}, is a chirp transient compatible with 
the emission from a binary neutron star system coalescence
in the late-inspiral phase \cite{TheLIGOScientific:2017qsa,Abbott:2018wiz,LIGOScientific:2018mvr}.
The signal has significant signal-to-noise ratio (SNR) in the range 30 to 600 Hz, roughly
corresponding to the last 100 to 30 orbits to merger for an
equal-mass binary with total mass $M\sim 2.7\:\Mo$.
The data analysis of GW170817 provided us with an estimate of the
dominant tidal polarizability parameter that, in turn, constrains the
NS cold equation of state \cite{Damour:2012yf,Favata:2013rwa,DelPozzo:2013ala,Lackey:2014fwa,Abbott:2018exr}.
The LIGO-Virgo detectors' sensitivity was not sufficient to detect a
signal from the merger phase and the remnant, which lie in the kHz range~\cite{Abbott:2018hgk}. An outstanding question is thus whether
the coalescence resulted in the formation of a black-hole (BH) or in a NS
remnant. 

A first answer was given by the interpretation of the electromagnetic
counterparts observed with delays of seconds to days with respect to
the GW and composed by a GRB \cite{Monitor:2017mdv,Troja:2017nqp,GBM:2017lvd} and a kilonova \cite{Coulter:2017wya,Chornock:2017sdf,Nicholl:2017ahq,Cowperthwaite:2017dyu,Tanvir:2017pws,Tanaka:2017qxj}.
Energetics and timing of the latter exclude both a prompt BH formation
and a long-lived remnant, e.g.~\cite{Margalit:2017dij,Shibata:2017xdx}.
Most likely, the merger dynamics produced a hypermassive NS that
collapsed on timescales of ${\sim}0.01$ to ${\sim}2$ seconds. 
Such a conclusion is informed and supported by numerical relativity
(NR) results that established the formation of hypermassive NS remnants for
canonical NS masses and equations of state supporting $\Mmax\gtrsim2\Mo$ 
\cite{Shibata:1999wm,Shibata:2003ga,Anderson:2007kz,Baiotti:2008ra,Thierfelder:2011yi,Hotokezaka:2011dh,Bauswein:2013jpa,Shibata:2017xdx,Radice:2018xqa}.

In this work, we explore a different approach to inferring the merger
remnant. Instead of considering the EM counterparts, we consider 
the pre-merger GW and infer binary parameters using
the late-inspiral solely. The posterior distributions of these
parameters are then combined with information from NR simulations. Our 
methods allow us to quantify the probability that a BH was promptly formed. 

This paper is structured as follows:
Sec.~\ref{sec:threshold} outlines the input from NR data in our inference methods, based on which we classify the outcome of a BNS merger;
the two methods are introduced in Sec.~\ref{sec:analysis} and are validated by analyzing a set of simulated GW detections in Sec.~\ref{sec:validation};
we perform the analysis on GW170817 data and present our results in Sec.~\ref{sec:GW170817}, while some concluding remarks are given in Sec.~\ref{sec:conc}.
We use geometric units $G=c=1$ unless stated differently.

\section{Prompt Collapse Threshold}
\label{sec:threshold}

\subsection{Mass threshold estimate}
\label{sec:Mthr_estimate}

\begin{figure}[t]
  \centering
    \includegraphics[width=0.49\textwidth]{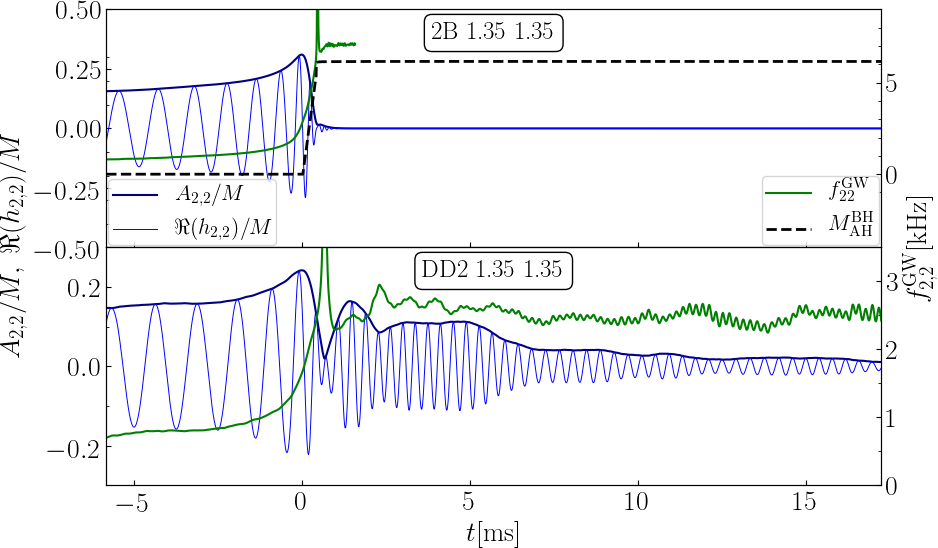}
    \caption{Example of waveforms for a binary neutron star merger for two different EOS
      but same component NS masses $M_1=M_2=1.35~\Msun$. Upper panel: 
      prompt collapse occurs after collision; the waveform amplitude drops to 0 while an apparent horizon (dashed black line) originates within 2~ms from the waveform peak amplitude. Bottom panel: a stable massive NS remnant 
      forms.}
    \label{fig:collapse_example}
\end{figure}

Numerical-relativity simulations indicate that a NS binary merger will
be followed by a prompt collapse to a BH, if the total gravitational mass $M$ of the
binary exceeds a threshold mass, given by \cite{Hotokezaka:2011dh,Bauswein:2013jpa} 
\be\label{eq:Mthr}
\Mthr = \kthr \Mmax \ .
\ee
In the expression above, $\kthr$ depends, in general, on the
EOS, mass ratio, and spin, 
while $\Mmax$ is the gravitational mass of the
heaviest stable nonrotating NS, which also depends on the EOS.
Empirically, the prompt collapse threshold is calculated from the
simulations by considering remnants that collapse
within 2~ms from the waveform peak amplitude 
(conventionally,  the ``merger time''). Examples of merger waveforms
for a prompt collapse and a NS remnant are shown in Fig.~\ref{fig:collapse_example}.
In the prompt collapse case an apparent horizon forms during the
simulation at a time close to the retarded merger time; the waveform frequency at
those times corresponds to the quasi-normal mode of the black hole.

For a sample of hadronic EOS and equal-mass nonspinning binaries,
the threshold coefficient in Eq.~\eqref{eq:Mthr} is found in the range
\cite{Hotokezaka:2011dh,Bauswein:2013jpa,Koppel:2019pys} 
\be\label{eq:k_thr}
1.3 \lesssim \kthr \lesssim 1.7 \ .
\ee
Considering a sample of equal-mass, nonspinning binaries and 12
hadronic EOS, Ref.~\cite{Bauswein:2013jpa} showed that $k_\text{trh}$
has an approximately EOS-independent linear
behavior in the maximum compactness $C_\text{max}$ of nonrotating
equilibrium NS solution.
Note that by inverting Eq.~\eqref{eq:Mthr} and {\it assuming} that
the merger did not promptly form a BH, one may obtain a bound on
the maximum stable NS mass~\cite{Shibata:2017xdx,Bauswein:2017vtn,Ruiz:2017due}.

We have repeated the analysis on the threshold mass with the data of
{\tt CoRe} collaboration \cite{Zappa:2017xba,Dietrich:2018phi} by
including 10 new simulations with 5 EOS and different
masses and spins.
We have compared and combined our new results with the ones reported
in \cite{Hotokezaka:2011dh,Bauswein:2013jpa,Koppel:2019pys}. Our final
sample includes 18 different hadronic EOS and for 8 of them results
from more than one analysis are available. 
Using the results reported by \cite{Bauswein:2013jpa} 
and \cite{Koppel:2019pys}, and by adding the data of
{\tt CoRe} collaboration, we find a linear fit with updated
coefficients that reads 
\begin{equation} \label{eq:k_fit}
  k_\text{thr}(C_\text{max}) = -(3.29\pm0.23) \, C_\text{max} + (2.392\pm0.064)  \ .
\end{equation}
The data that were used for this fit are given in Appendix~\ref{app:pckthr}, along
with further details.

\subsection{Tidal parameter threshold estimate}
\label{sec:LambdaT_estimate}

Alternatively, the prompt collapse threshold
can be characterized in an EOS-independent way in terms of the
tidal polarizability parameter 
\begin{align}\label{eq:kappaT2}
  \kappa_2^\text{T} &= \dfrac{3}{2} \left[ \Lambda_2^\text{A} X_\text{A}^4 X_\text{B} + \Lambda_2^\text{B} X_\text{B}^4 X_\text{A} \right] \ ,
\end{align}
where the tidal polarizability coefficient of star $A$ is
\be
\label{eq:def_Lam2}
\Lambda^\text{A}_2 = \dfrac{2}{3} k^\text{A}_{2} \left(\dfrac{c^2}{G}
\dfrac{R_\text{A}}{M_\text{A}}\right)^5 \ ,
\ee
and $k^\text{A}_{2}$ is the quadrupolar gravito-electric Love number~\cite{Hinderer:2007mb,Damour:2009vw,Binnington:2009bb}.
Above, $(R_\text{A},M_\text{A})$ are the NS areal radius and
mass and $X_\text{A}=M_\text{A}/M$.
The $\Lambda_2$ parameter is strongly dependent on the
NS internal structure; thus, its measurement provides a constraint
on the NS EOS~\footnote{Black holes are not deformed in this way;
  black hole static perturbations lead to
  $k_2=0$~\cite{Damour:2009va,Damour:2009wj,Binnington:2009bb,Gurlebeck:2015xpa}.}.
The binary's post-Newtonian tidal dynamics and waveform are parametrized at leading-order 
by $\kappa^\text{T}_2$ \cite{Damour:2012yf,Bernuzzi:2014kca}.
A tidal polarizability parameter commonly used in GW analysis 
(and equivalent to $\kappa^\text{T}_2$ for equal-mass binaries) is 
\begin{align}
  \label{eq:LambdaT}
  \tilde\Lambda &= \frac{16}{13}
  \frac{(M_\text{A} + 12 M_\text{B}) M_\text{A}^4}{M^5}\Lambda_\text{A} + (A\leftrightarrow B)\ .
\end{align}

By analyzing the NR data of the {\tt CoRe} collaboration,
Ref.~\cite{Zappa:2017xba} found that all the reported prompt collapse mergers are captured by the condition
$\kappa^\text{T}_2 < 73$,
with a variability of $\delta \kappa^\text{T}_2 < 40$, depending on the EOS. 
%
%
%
Inspection of the same NR data provided also a range for the value of $\tilde\Lam$ at the prompt collapse threshold:
\be\label{eq:Lam_thr}
338 \lesssim \tilde\Lam_\text{thr} \lesssim 386 \ .
\ee


\section{Method}
\label{sec:analysis}

Based on the universal behavior discussed in Sec.~\ref{sec:threshold},
we present two different ways of inferring whether a BNS merger is
followed by a prompt collapse to a BH using solely GW data (with the
exception of the sky location which we may fix to the one obtained by
EM observations, when an EM counterpart is available).
We test the validity of our methods against a set of high-resolution
numerical simulations of BNS mergers with different masses and EOS.

For our Bayesian data analysis on the GW signal, we use a Markov-chain
Monte Carlo (MCMC) algorithm as implemented in the LALInference
software package~\cite{Veitch:2014wba}, with a set-up similar to the one
employed in the latest LVC analysis of
GW170817~\cite{Abbott:2018wiz,Abbott:2018exr}. 

\subsection{Threshold Mass}
\label{sec:mthr}

For this method we make use of the mass threshold estimate of Sec.~\ref{sec:Mthr_estimate},
whereby the total mass $M$ of the progenitor NS binary being larger or
smaller than $\Mthr$ determines whether
the product of the merger will promptly collapse to a BH or not.
The threshold mass $\Mthr$ depends on the EOS via Eq.~\eqref{eq:Mthr} and~\eqref{eq:k_fit}.
We perform a full Bayesian analysis on the data, that returns posterior distributions
for the binary parameters, including the EOS. The barotropic EOS for the cold dense NS matter is sampled through a 4-dimensional family of
pressure-density functions $P(\rho)$, parametrized by $(\gamma_{0},\gamma_{1},\gamma_{2},\gamma_{3})$
in the spectral
decomposition~\cite{Lindblom:2010bb,Carney:2018sdv,Abbott:2018exr}, a
smooth alternative to piecewise polytropic models
\cite{Read:2008iy,Lackey:2014fwa}, where the adiabatic index $\Gamma =
\rho \frac{\dee\ln P}{\dee\rho}$ 
is given by
\begin{equation}
  \label{eq:Gamma_spec}
  \Gamma =  \exp
  \left[
    \sum_{k=0}^{3} \gamma_{k} \log (p/p_{0})^{k}
  \right] \, .
\end{equation}
with $p_{0}$ some reference pressure.
For each sampled point in the parameter space, we solve the TOV equations to calculate not only 
the tidal polarizability parameters $\Lambda_{2}^{A}$ which are used to model the tidal effects in
the waveform, but also the values for $\Mmax$, $C_{\rm max}$ and $\kthr$.
We can thus translate the joint posterior PDF on masses and EOS parameters $(m_{1}, m_{2}, \gamma_{0}, \gamma_{1},\gamma_{2},\gamma_{3})$
into a joint posterior PDF on the $(M,\Mthr)$ plane. The
fraction of the posterior distribution that 
lies above the diagonal is equal to the posterior probability of prompt collapse
\begin{equation}
  \label{eq:Ppc_Mthr}
  P_{\rm PC} = P(M > \Mthr(\vec\gamma) | d) \, ,
\end{equation}
where $\vec{\gamma} = (\gamma_{0},\gamma_{1},\gamma_{2},\gamma_{3})$ and $d$ denotes our data. 


As an additional step, one may choose to impose further implicit constraints on the parameter space,
such as requiring that the EOS support NS masses larger than a given value. For instance, the
observation of the binary pulsar PSR J0348+0432~\cite{Antoniadis:2013pzd} and the narrow
measurement of the pulsar's mass, or even the more recent measurement of 
an even heavier (but with larger uncertainty) pulsar mass in J0740+6620~\cite{Cromartie:2019kug}.
In one of the analyses of~\cite{Abbott:2018exr} the conservative 1-$\sigma$ bound for
PSR J0348+0432 at $1.97 \Mo$ was considered as a hard constraint.
Here we take a different approach and marginalize over the mass measurement uncertainties into our analysis by
treating that measurement as a random variable sampled from a normal
distribution that is adapted to the mean and standard deviation of the
measurement, and weighing the posterior samples accordingly. 


\subsection{Threshold Tidal Parameter}
\label{sec:lambdat}

For the second method we again employ a Bayesian analysis of the
GW data, this time focusing on the posterior distribution of the tidal
deformability parameter $\tilde{\Lambda}$ given by
Eq.~\eqref{eq:LambdaT}. 
The set-up of our Bayesian analysis follows that of~\cite{Abbott:2018wiz}.
We then make use of the corresponding criterion of
Eq.~\eqref{eq:Lam_thr} in order to estimate the probability of prompt
collapse. 
Note that the criterion defines a transition region between the
studied cases where the merger product undergoes prompt collapse and
the ones where it does not. 
The outcomes of NR simulations within this transition region are not
perfectly ordered. 
We treat this classification problem by assigning a probability
distribution to the uncertainty of the threshold value
$\tilde\Lam_\text{thr}$ instead of choosing a hard threshold or,
equivalently, by defining a sigmoid-type conditional probability of
prompt collapse for a given value of $\tilde{\Lambda}$ as 
\begin{equation}
  \label{eq:prob_Lthr}
  P(\text{prompt collapse}|\tilde{\Lambda}) = \frac{1}{1 +
    e^{\frac{\tilde\Lambda - \tilde{\Lambda}_{0}}{\beta}}} \, , 
\end{equation}
which tends to $1$ ($0$) for small (large) values of $\tilde{\Lambda}$.
The values for the sigmoid parameters, i.e. the central value and the
width, are chosen based on the available set of NR simulations in this
region to be $\tilde{\Lambda}_{0} \approx 362$ and $\beta \approx
13.7$ respectively.   
Then, once the posterior PDF $p(\tilde{\Lambda}|d)$ is calculated, the
probability of prompt collapse is simply 
computed by integrating the posteriors from the minimum value up
to the threshold value using the sigmoid of Eq.~\eqref{eq:prob_Lthr}
as a kernel 
\begin{equation}
  \label{eq:Ppc_LambdaT}
  P_{\rm PC} = \int \dee \tilde{\Lambda} \, P(\text{prompt
    collapse}|\tilde{\Lambda}) \, p(\tilde{\Lambda}|d) \, . 
\end{equation}
Note that this method does not rely on any assumption about the EOS,
but only on the phenomenological parameter $\tilde{\Lambda}$ which is
directly measured from the data. 
In the present analysis we assume that Eq.~\eqref{eq:Lam_thr}
holds independently on $q$ and spins. That hypothesis is justified
by inspection of the {\tt CoRe} data that span $q\in[1,2]$ and
dimensionless spins up to $\sim0.1$.


\begin{table*}[t!]
  \centering    
  \caption{Summary of injections and {\TF2} recovery.
    Collapse time $t_\text{BH}$ is reported from merger time, defined at the peak of
    the amplitude. We indicate with HMNS (MNS) remnants that are short
    (long) lived, i.e. that (do not) collapse to BH within the simulated time.
    All the simulations are performed at standard resolution of
    \cite{Radice:2018pdn}.}

  \resizebox{\textwidth}{!}{
\begin{tabular}{cccccccccccccccccc}        
    \hline
    \hline
    EOS & $\Mmax$  & $C_\text{max}$  & $\Mthr$  &
    $M_A$ & $M_B$  & $\mathcal{M}_{c}$ & $\Lam^A_2$ & $\Lam^B_2$ & $\tilde\Lam$ &
    $t_\text{BH}$& Remnant at &
    Ref. & $\PPC^{\Mthr}$ & $\PPC^{\Mthr,\Mmax}$ &  $\PPC^{\tilde{\Lambda}_{\rm thr}}$
    \\
         &  $[\Mo]$& & $[\Mo] $&
   $[\Mo]$ &$[\Mo]$  &$[\Mo]$ && & &
     [ms] &  $t\sim3$~ms &  & \% & \% & \%\\
    \hline
    2B & $1.78^*$ & 0.3120 & 2.43$^{+0.24}_{-0.24}$  & 1.35 & 1.35  & 1.17 & 127 & 127 & 127 & 0.49 & BH & \cite{Bernuzzi:2014owa} & $99.5$ & $48.3^*$ & $100$\\
    SLy & 2.06&0.3066  & 2.87$^{+0.26}_{-0.30}$ & 1.50 & 1.50 & 1.30 & 191 & 191 & 191 & 0.99 & BH & \cite{Dietrich:2018phi}  & $94.5$ & $82.9$ & $100$\\ 
    LS220 &2.04 &0.2841 &2.95$^{+0.29}_{-0.24}$  & 1.60 & 1.60  &1.39& 202 & 202 & 202 & 0.63 & BH & \cite{Radice:2017lry,Radice:2018xqa,Radice:2018pdn} & $90.0$ & $84.4$ & $99.9$\\
    SFHo & 2.06 & 0.2952& 2.95$^{+0.25}_{-0.29}$ & 1.46 & 1.46  &1.27& 252 & 252 & 252 & 0.70 & BH & \cite{Radice:2017lry,Radice:2018xqa,Radice:2018pdn} & $72.7$ & $40.4$ & $97.8$\\
    BHB$\Lambda\phi$ & 2.11& 0.2677  & 3.10$^{+0.35}_{-0.18}$& 1.60 & 1.60  &1.39& 306 &
    306 & 306 & 0.99 & BH & \cite{Radice:2016rys,Radice:2017lry,Radice:2018xqa,Radice:2018pdn} & $36.0$ & $19.6$ & $72.9$\\
    DD2  & 2.42& 0.3007  & 3.35$^{+0.37}_{-0.28}$  &  1.59 &  1.59 & 1.38 & 332 & 332 & 332  &  $\sim3$ & BH & \cite{Radice:2016rys} & $32.1$ & $18.5$ & $66.2$\\
    SFHo & 2.06 & 0.2952&  2.95$^{+0.25}_{-0.29}$  &1.40 & 1.40 &1.22& 334 & 334 & 334 & 1.07 & BH & \cite{Radice:2017lry,Radice:2018xqa} & $41.7$ & $4.9$ & $60.6$\\
    ALF2 & 1.99 & 0.2602 & 2.87$^{+0.43}_{-0.06}$ & 1.50 & 1.50 & 1.30 & 382 & 382 & 382 & 0.64 & BH & \cite{Dietrich:2018phi} & $20.9$ & $3.6$ & $29.2$\\
    SLy-SOR  & 2.06 & 0.3066   & 2.87$^{+0.26}_{-0.30}$   & 1.34 & 1.34 &  1.17 & 401 & 401 & 401 & $\sim14$ & HMNS & This work & $25.0$ & $0.4$ & $21.8$\\
    SLy-SOR & 2.06 &  0.3066   & 2.87$^{+0.26}_{-0.30}$  & 1.43 & 1.26  & 1.17 & 264 &  592&401 & $\sim13$ & HMNS & This work & $23.7$ & $0.1$ & $21.6$\\
    SFHo & 2.06  &0.2952 &2.95$^{+0.25}_{-0.29}$ & 1.44 & 1.27  &1.18 & 274 & 606 & 412 & $\sim12$ & HMNS & This work & $21.6$ & $2.0$ & $17$\\
    SFHo & 2.06 & 0.2952&2.95$^{+0.25}_{-0.29}$& 1.35 & 1.35  &1.18 &413 &413&413 & $\sim4$ & HMNS & \cite{Radice:2017lry,Radice:2018xqa,Radice:2018pdn} & $20.5$ & $0.2$ & $15.6$\\
    LS220 & 2.04 &0.2841 & 2.95$^{+0.29}_{-0.24}$ & 1.44 & 1.25  &1.17& 432 &1136 &713 & $\sim33$ & HMNS & This work  & $2.4$ & $0.0$ & $0$\\
    LS220 & 2.04 & 0.2841& 2.95$^{+0.29}_{-0.24}$  & 1.34 & 1.34  &1.17& 715& 715& 715& $\sim16$ & HMNS & This work  & $0.7$ & $0.0$ & $0$\\
    DD2 & 2.42& 0.3007& 3.35$^{+0.37}_{-0.28}$   &1.36 & 1.36 &1.18& 840& 840& 840& $\sim21$ & MNS & \cite{Perego:2019adq} & $0.1$ & $0.0$ & $0$\\
    DD2  & 2.42& 0.3007  & 3.35$^{+0.37}_{-0.28}$    &  1.24 &  1.24 & 1.08 & 1366 & 1366 & 1366 & $> 20$ & MNS & \cite{Radice:2018xqa} & $0.0$ & $0.0$ & $0$\\
    BHB$\Lambda\phi$  & 2.11 & 0.2677   & 3.10$^{+0.35}_{-0.18}$ &  1.24 &  1.24 & 1.08 & 1367 & 1367 & 1367 &  $> 20$ & MNS & \cite{Radice:2016rys} & $0.0$ & $0.0$ & $0$\\
    \hline
    \hline
  \end{tabular}
  }
  \label{tab:NRSim}
  \\
\end{table*}

\section{Injection studies}
\label{sec:validation}

We validate our methods using injections of known inspiral-merger
waveforms corresponding to binaries simulated in NR. We demonstrate
that both methods are effective in estimating the collapse treshold and
discuss/quantify their systematics.


\subsection{Setup}
\label{sec:simsetup}

We consider NR merger simulations of irrotational binaries with
different chirp masses specifically performed for this work
together with data previously presented in \cite{Bernuzzi:2014owa,Radice:2016rys,Radice:2017lry,Dietrich:2018phi,Radice:2018xqa,Radice:2018pdn,Perego:2019adq}.
The new simulations are performed with the \texttt{WhiskyTHC} code
\cite{Radice:2012cu, Radice:2013hxh, Radice:2013xpa} 
at multiple grid resolutions, using the same setup described in
\cite{Radice:2017lry,Radice:2018xqa,Radice:2018pdn}.

The main properties of the simulated binaries, the outcome of the
merger and the summary data from the injection are summarized in Tab.~\ref{tab:NRSim}. 
We simulate with five microphysical EOS:
the BHB$\Lam\phi$ EOS \cite{Banik:2014qja}, 
the DD2 EOS \cite{Typel:2009sy, Hempel:2009mc},
the LS220 EOS \cite{Lattimer:1991nc},
the SFHo EOS \cite{Steiner:2012rk},
the SLy-SOR EOS \cite{daSilvaSchneider:2017jpg}; and tthree piecewise
polytropic: the ALF2, 2B and the SLy EOS \cite{Read:2008iy}.
The microphysical EOS predict NS maximum masses and radii within the
range allowed by current astrophysical constraints.
The 2B EOS is representative of soft EOS that do not support the largest NS masses
observed so far~\cite{Antoniadis:2013pzd,Cromartie:2019kug}.
All the simulations with chirp mass $\M_c\sim 1.18\:\Mo$ and
tidal deformability compatible with GW170817 except 2B, predict a short-lived NS
remnant collapsing to BH within ${\sim}15$~ms.
The simulation DD2 1.59+1.59 with $\tilde\Lambda=332$ is below the
$\tilde\Lambda$ threshold for prompt BH formation
but forms a short lived NS with lifetime ${\sim}3$~ms
\cite{Radice:2016rys}. While this is possibly related to 
numerical uncertainties, the binary provides an interest borderline case.

Throughout this work we use the terms hypermassive NS (HMNS) and 
massive NS (MNS) with a slightly different meaning than what is
commonly used in the literature~\footnote{%
A HMNS is defined as a differentially rotating NS at equilbrium with mass
above the uniformly rotating limit \cite{Baumgarte:1999cq}. A
supramassive NS is a rotating NS at equilibrium with rest mass
exceeding the nonrotating limit $\Mmax$ \cite{Cook:1993qr}. A remnant
with mass below $\Mmax$ is usually indicated as MNS.}
We indicate with HMNS (MNS) merger remnants that are short (long)
lived, i.e. that (do not) collapse to BH within the simulated time. 
The reason for this choice is that merger remnants are not cold
equilibrium NS configurations, and their secular evolution is far from
being understood (see e.g. discussion in \cite{Radice:2018xqa}.)

The simulations
provide us with dynamics and waveform starting from GW frequencies
$\sim\,500-900$~Hz, depending on the binary mass and simulation length. Hence, the NR waveform alone are not sufficient to perform injection of BNS signals.
Waveforms spanning from an initial GW frequency of 30~Hz to
merger and corresponding to the binaries of Tab.~\ref{tab:NRSim} are constructed 
with the {\tt TEOBResumS} 
waveform model \cite{Bernuzzi:2014owa,Nagar:2018zoe}. Specifically, we use the
nonspinning tidal model of \cite{Akcay:2018yyh} with gravitational-self-force resummed 
gravitoelectric and post-Newtonian gravitomagnetic terms
(Model GSF23$^{(+)}$PN$^{(-)}$ with $p=4$ of Tab.~I in
\cite{Akcay:2018yyh}). Waveforms are generated using 
the post-adiabatic inspiral speed-up developed in \cite{Nagar:2018gnk}~\footnote{The public available {\tt TEOBResumS} code can be found at \cite{teobresums}.
}.

For our Bayesian data analysis on the simulated GW signals, we use a Markov-chain
Monte Carlo (MCMC) algorithm as implemented in the LALInference
software package~\cite{Veitch:2014wba,lalsuite}, with a set-up similar to the one
employed in the latest LVC analysis of GW170817~\cite{Abbott:2018wiz,Abbott:2018exr}.
The simulated signals are coherently projected and analyzed as the output strain of
LIGO Handford (H1), LIGO Livingston (L1) and Virgo (V1) at design sensitivity.
The intrinsic parameters of the nonspinning BNS sources are given in Tab.~\ref{tab:NRSim},
while the location and orientation parameters are compatible with GW170817.
In order to isolate possible systematics from statistical uncertainties due to noise,
we perform our tests on the zero-noise realization.

\begin{figure*}[t]
  \centering
  \includegraphics[width=0.52\textwidth]{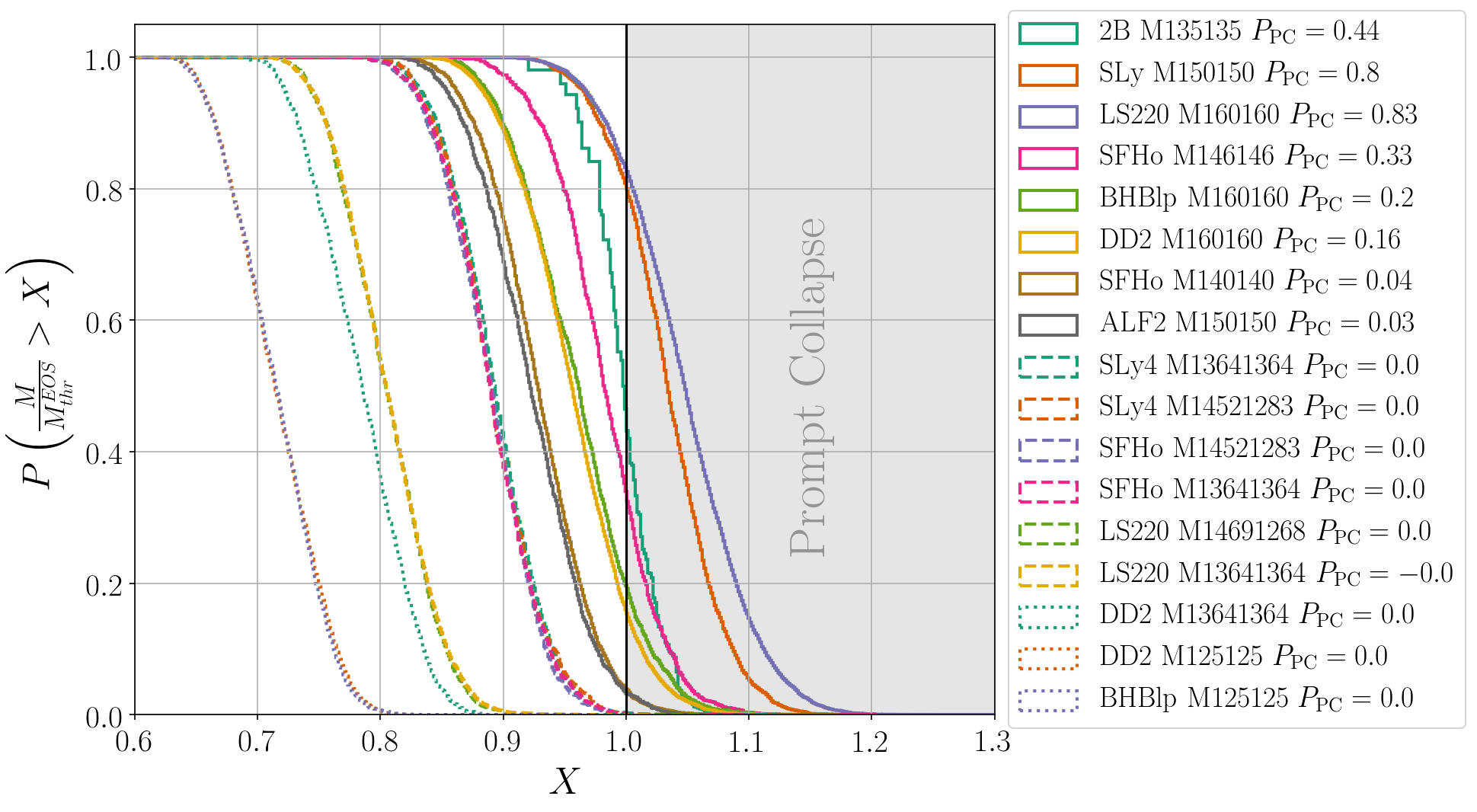}
  \includegraphics[width=0.46\textwidth]{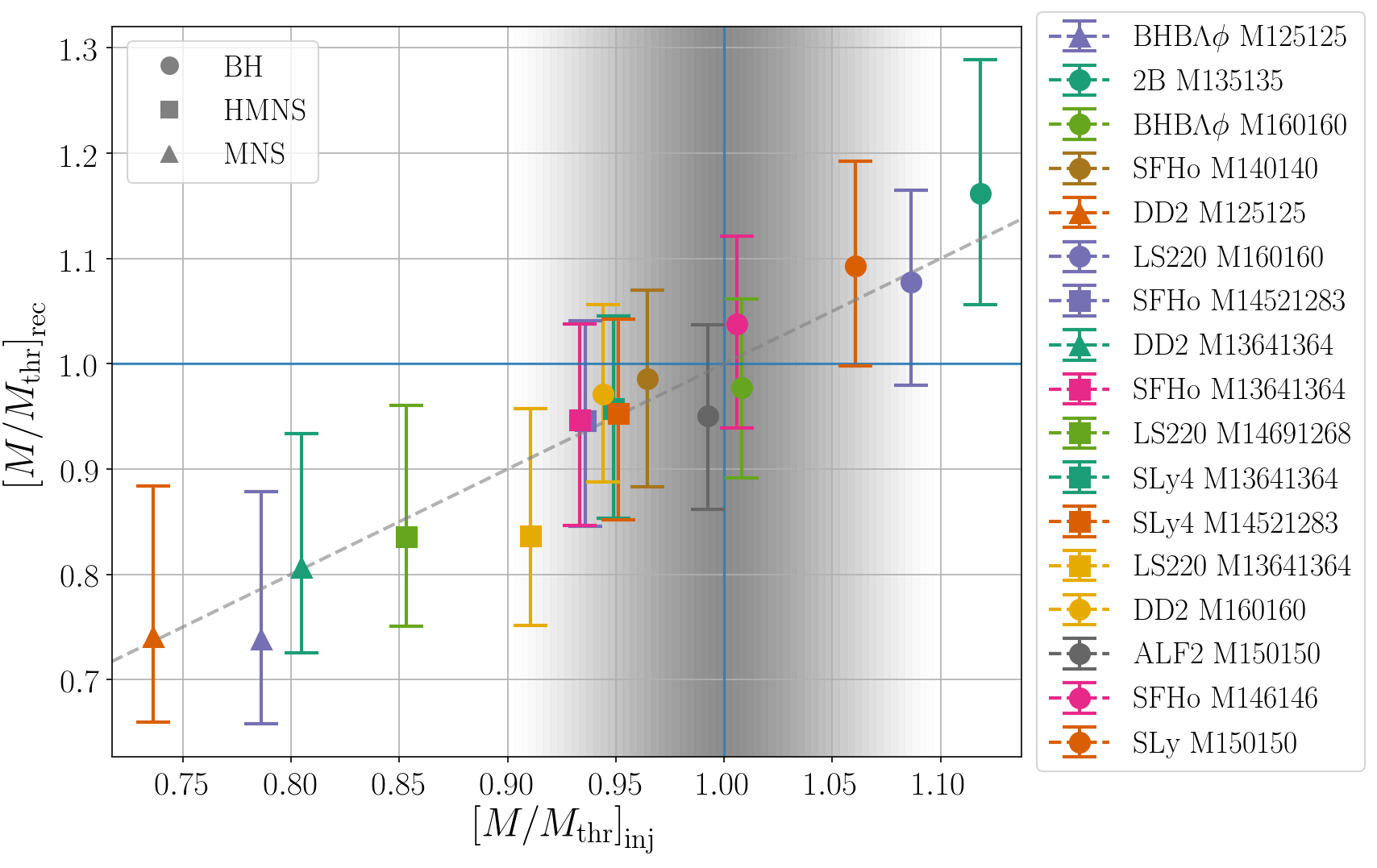}
  \caption{Results of injection study, for the threshold-mass method using the simulated BNS events of Tab.~\ref{tab:NRSim}.
    Left: Cumulative probability of $M/\Mthr$, the ratio between total mass and threshold mass.
    The inferred probability of prompt collapse for each BNS event is the value of its curve at $X=1$.
    Solid, dashed and dotted lines indicate a BH, HMNS and MNS remnant respectively.
    Right: Summary of the injected values of $M/\Mthr$ vs the recovered median values and $90\%$ confidence intervals, for the simulated BNS events.
    Circles, squares and triangles indicate a BH, HMNS and MNS remnant respectively (from NR).
    The threshold uncertainty due to the error in the fitting formula of Eq.~\eqref{eq:k_fit} is shown as the grey shaded band.}
    \label{fig:inj:mmax}
\end{figure*}

\begin{figure*}[t]
  \centering
  \includegraphics[width=0.52\textwidth]{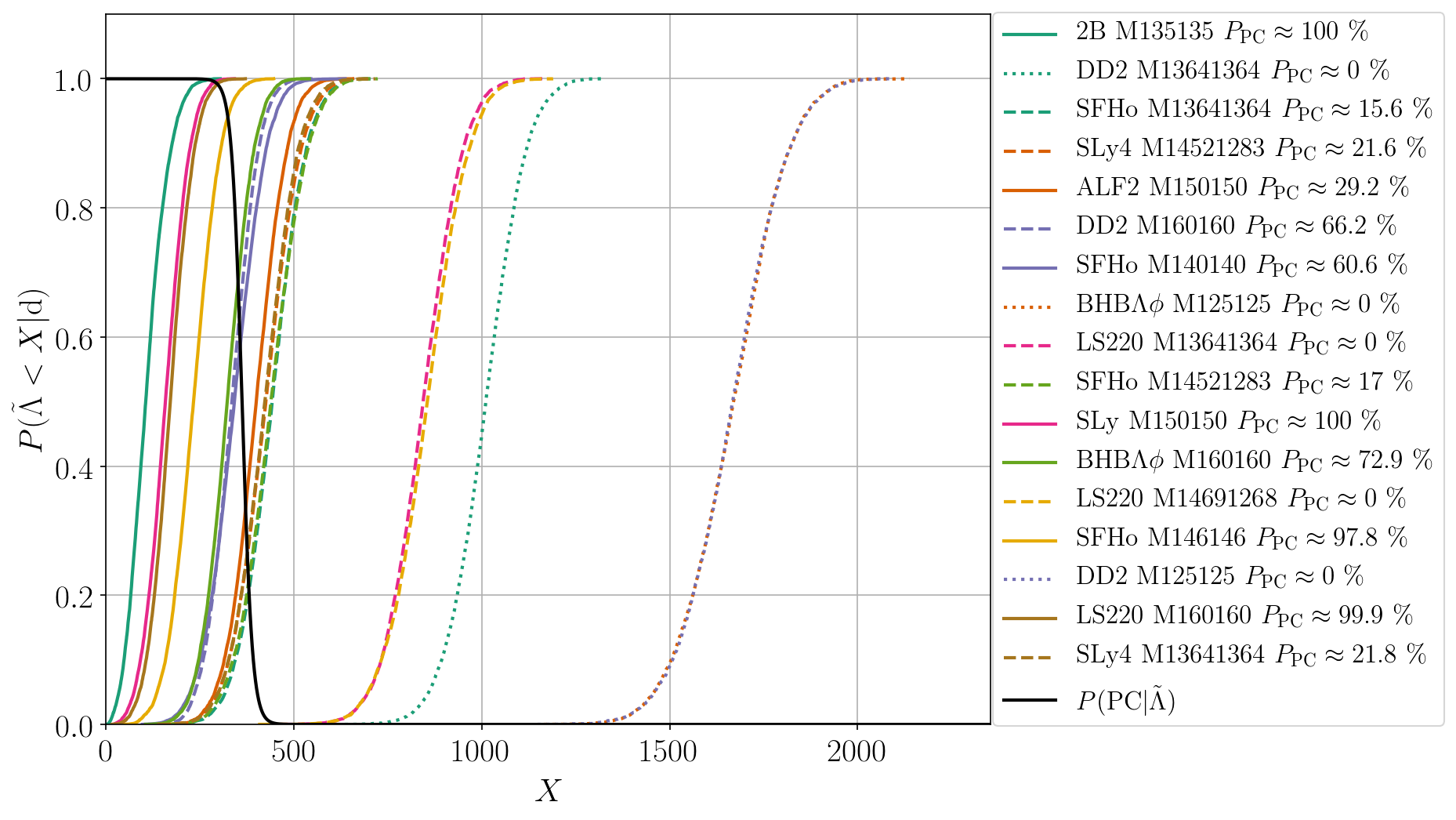}   
  \includegraphics[width=0.46\textwidth]{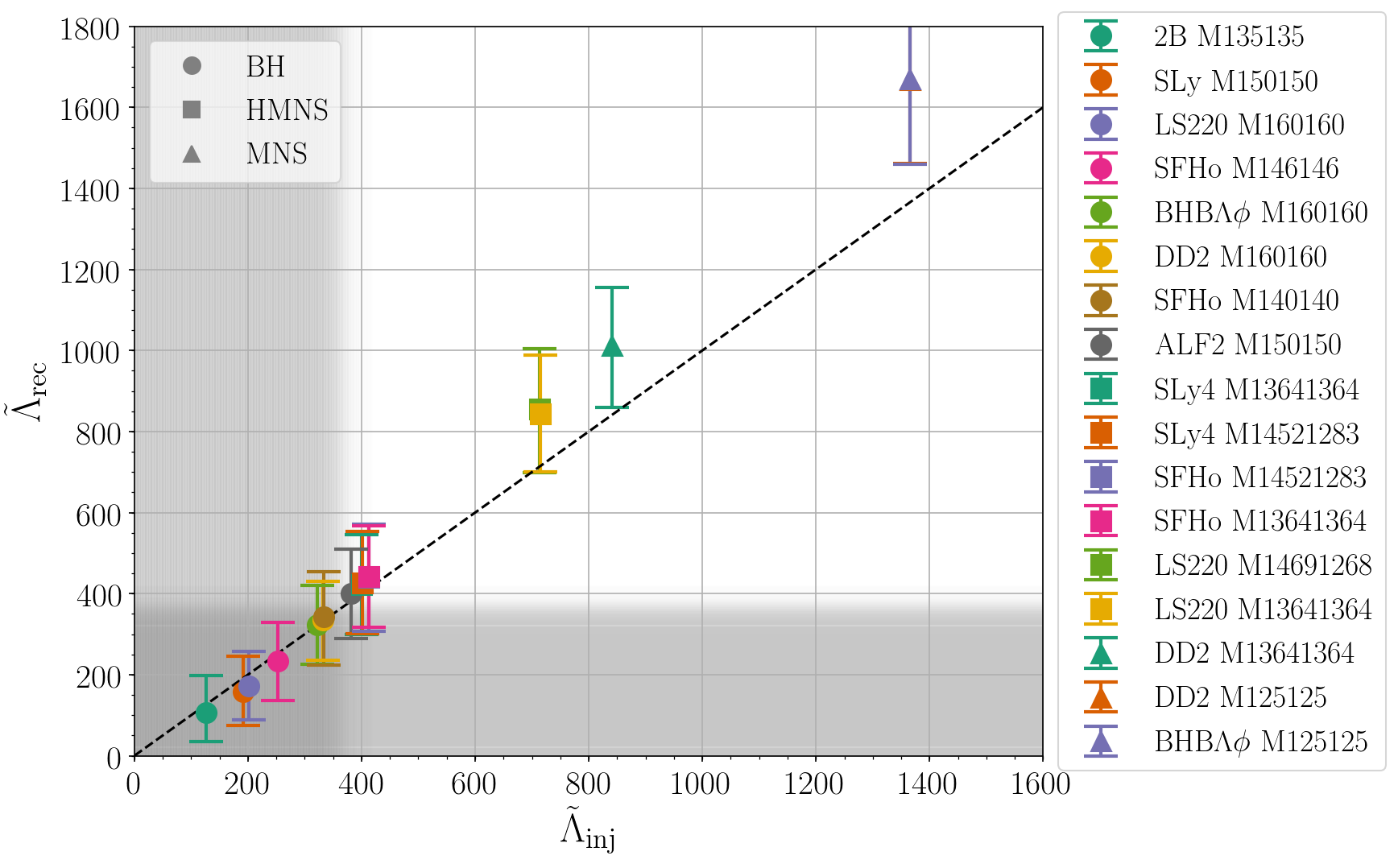}
  \caption{Results of injection study, for the threshold-$\tilde{\Lambda}$ method using the simulated BNS events of Tab.~\ref{tab:NRSim}.
    Left: Cumulative probability of $\tilde{\Lambda}$ and the inferred probability of prompt collapse for each event.
    Solid, dashed and dotted lines indicate a BH, HMNS and MNS remnant respectively.
    The solid black line corresponds to the prior probability of prompt collapse as a function of $\tilde{\Lambda}$ (Eq.~\eqref{eq:prob_Lthr}).
    Right: Summary of injected values of $\tilde{\Lambda}$ vs recovered median values and $90\%$ confidence intervals, for the simulated BNS events.
    Circles, squares and triangles indicate a BH, HMNS and MNS remnant respectively (from NR).
  }
    \label{fig:inj:lamb}
\end{figure*}

We perform our analyses using two different waveform models, namely
{\TF2} and {\IMRP}, both restricted to aligned dimensionless
spins ranging within $[-0.05, 0.05]$ (the low-spin prior
of~\cite{Abbott:2018wiz}). 
The tidal effects are modeled in the case of {\TF2} up to 2.5PN
beyond leading order~\cite{Damour:2012yf} and in the case of
{\tt IMRPhenomPv2} using the {\tt NRTidal} approach of~\cite{Dietrich:2017aum}. 

In the threshold-mass method, we are able to additionally impose an observational constraint on
the EOS prior, coming from the heaviest observed NS. This can be either a hard constraint at a
chosen mass value (e.g. $1.97$~$\Mo$ as in~\cite{Abbott:2018exr}), or a probabilistic
constraint that takes into account the measured posterior PDF.
In the latter case we will use the median and
1-$\sigma$ error of the mass measurement of PSR J0348+0432~\cite{Antoniadis:2013pzd}
to reconstruct a Gaussian PDF for the heaviest observed NS mass 
$p(\overline{M_{\rm max}}) = \mathcal{N}(2.01 \Mo, 0.04 \Mo)$
and assign a weight $w$ on each sampled EOS appropriately, according to its
maximum NS mass,
\begin{equation}
  \label{eq:weights}
  w(\vec\gamma) = p(\Mmax(\vec\gamma) > \overline{M_{\rm max}}) \, .
\end{equation}
A comparison between results derived with and without such a constraint is demonstrated
in Appendix~\ref{app:maxM}.

\subsection{Results}
\label{sec:simresults}

We find that the {\IMRP} waveform systematically underestimates the inference
of the injected $\tilde\Lambda$. This result was anticipated by the high SNR
injections of \cite{Dudi:2018jzn,Messina:2019uby}, but could not be
studied systematically due to the limited number of injections
performed there.
A similar bias is present in the EOS inference runs with {\IMRP}, but the mass
threshold method to determine the prompt collapse is less affected by
systematics on tidal parameters than the $\tilde\Lambda$ threshold method.  
In the following, we discuss only the results obtained with
{\TF2}. The effect of waveform systematics on the results is discussed 
in Appendix~\ref{app:WfSys}; a full account of the waveform's
systematics in these injection experiments will be given elsewhere [In Prep.].

The recovery results with {\TF2} are summarized in Tab.~\ref{tab:NRSim}.
Results from the threshold-mass analysis with maximum mass constrained
to be larger than the mass of PSR J0348+0432 are also shown in
Fig.~\ref{fig:inj:mmax} [See Appendix~\ref{app:maxM} for a similar plot
  without the maximum mass constraint]. The left panel shows for each injection the
cumulative probability distribution to find $M>\Mthr$; the
vertical line marks the collapse threshold. The right panel shows for each
injection the inferred mass divided by the inferred threshold mass 
$\left[M/\Mthr\right]_{\rm rec}$ versus the same injected quantity,
$\left[M/\Mthr\right]_{\rm inj}$. Erroneous recoveries would populate the top-left and
bottom-right quadrants of the plot. The plot shows that the inference
returns the correct 
estimate of the prompt collapse for the majority of the injections.
SFHo 1.46+1.46 is a borderline case for which
$P_\text{PC}\sim40\%$.
However, we observe that a few simulations that led to a prompt collapse (denoted 
by solid lines and circles in the plots), were not recovered as such.
This misclassification had occured already at the level of injection parameters,
since their position on the x-axis (right panel) lies below unity, which is
due to the inherent uncertainty on the estimation of $\kthr$ from fitting NR data.
The resulting error is comparable in size to the statistical error of our parameter estimation.

Results from the threshold tidal parameter analysis are
shown in Fig.~\ref{fig:inj:lamb}. The left panel shows for each
injection the probability that $\tilde\Lambda$ is smaller than a given
value. The latter should be compared to Eq.~\eqref{eq:prob_Lthr},
shown as a black solid line. The right panel summarizes the inference
results in a way analogous to the Fig.~\ref{fig:inj:mmax}.
The threshold tidal parameter analysis incorrectly predicts few
cases for BNS with $\tilde\Lambda\sim\tilde\Lambda_\text{thr}$.  
SFHo 1.40+1.40 ($\tilde\Lambda=334$) and ALF2 1.50+1.50 ($\tilde\Lambda=382$)
are predicted with 61\% and 29\% probability of producing a NS remnant
while the simulations indicate prompt BH formation.
The two SLy-SOR binaries ($\tilde\Lambda\sim401$) are predicted with $\sim22$\% probability of
prompt collapse while the merger produces a HMNS. 
DD2 1.59+1.59 ($\tilde\Lambda=332$) is predicted to promplty form a BH
with 66\% probability, with the simulation resulting in a HMNS of very
short life of ${\sim}3$~ms.

We conclude that both methods infer correctly the merger outcome of the simulations,
except for few cases corresponding to BNS close to the collapse threshold.   
Excluding these cases (in which the answer given is anyway inconclusive),
the mass threshold analysis with the maximum mass constraint better
captures the formation of a NS remnant, while the threshold
$\tilde\Lambda$ analysis captures more robustly the prompt collapse
cases.
The mass threshold analysis without the maximum mass constraint give
instead results comparable to the threshold
$\tilde\Lambda$ analysis. This can be understood by inspecting
the EOS posteriors in Appendix~\ref{app:EOSrec}. The EOS inference from the
inspiral data constrains more strongly the EOS at densities comparable
to the maximum density of the individual NS in the binary. These 
densities are those that determine the individual masses and thus the
$\Lambda_A$ parameters. Introducing a maximum mass constraint effectively
corresponds to introduce a lower bound on the mass distribution of the
individual NS (and on $\kthr$), thus lowering the collapse
probability. 

\section{Application to GW170817}
\label{sec:GW170817}

We apply our analysis methods to data from the first detected BNS event
GW170817, by postprocessing the publicly available posteriors of the
LIGO-Virgo collaboration released with \cite{Abbott:2018exr,Abbott:2018wiz}.
In all of the analysis set-ups, the NS spins are aligned with
the orbital angular momentum and the spin magnitudes are restricted to the
``low-spin'' prior range $\chi_{1,2} \in [-0.05,0.05]$.


\subsection{Results}
\label{sec:GW170817_results}

For the threshold-mass method (Sec.~\ref{sec:mthr}), we process the 
posteriors of the spectral parameters $(\gamma_{0},\gamma_{1},\gamma_{2},\gamma_{3})$
published in~\cite{Abbott:2018exr},
\begin{itemize}
\item without imposing an implicit constraint on $\Mmax$;
\item with the additional hard cut of $\Mmax \ge 1.97 \Mo$, corresponding to a $1\sigma$ conservative estimate of
   the PSR J0348+0432 mass measurement $2.01 \pm 0.04 \Mo$~\cite{Antoniadis:2013pzd};
\item with the additional probabilistic weight quantifying the probability of $\Mmax$ being heavier
  than the PSR J0348+0432 mass. 
\end{itemize}

For the threshold-$\tilde\Lambda$ method, we process published posteriors
on tidal parameters from a number of different analyses.
In particular, we consider methods that extend the BBH parameter space by
the matter-related parameters
\begin{itemize}
\item $(\tilde{\Lambda},\delta\tilde{\Lambda})$, using four different
  waveform models ({\IMRP}, {\IMRD}, {\SEOB}, {\TF2})
\item $\Lambda_{s} = (\Lambda_{1}+\Lambda_{2})/2$, the symmetric tidal 
parameter, using {\IMRP}, along with the use of the EOS-insensitive relation
for the antisymmetric tidal parameter $\Lambda_{a}(\Lambda_{s},q)$
(see~\cite{Abbott:2018exr}), which can then be mapped to $\tilde{\Lambda}$;
\item $\vec\gamma = (\gamma_{0},\gamma_{1},\gamma_{2},\gamma_{3})$ parametrizing,
  the EOS, from   which $\tilde{\Lambda}$ can be derived, with and without a constraint
  on $\Mmax$ (see parametrized EOS method 
of~\cite{Abbott:2018exr}).
\end{itemize}

\begin{figure*}[t]
  \centering 
    \includegraphics[width=0.49\textwidth]{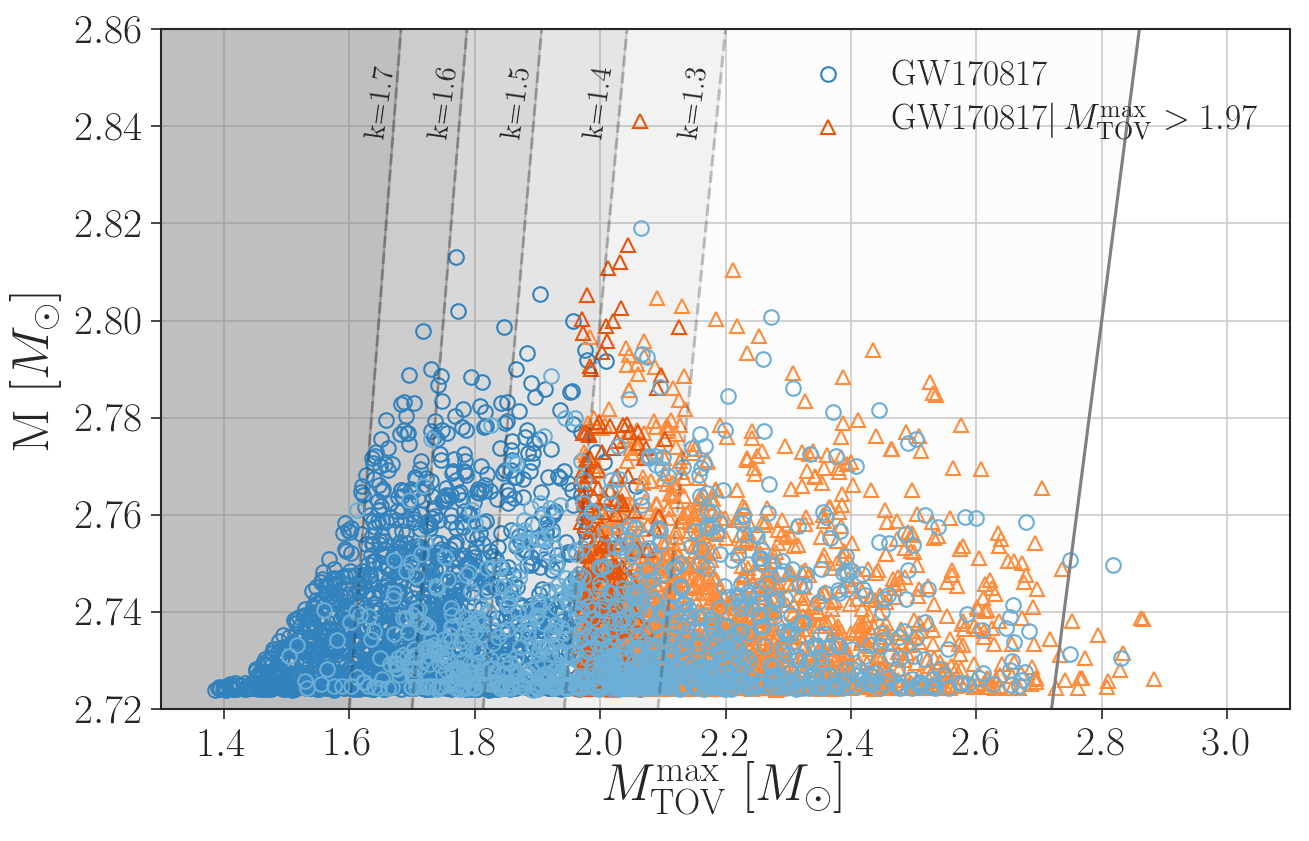}
    \includegraphics[width=0.49\textwidth]{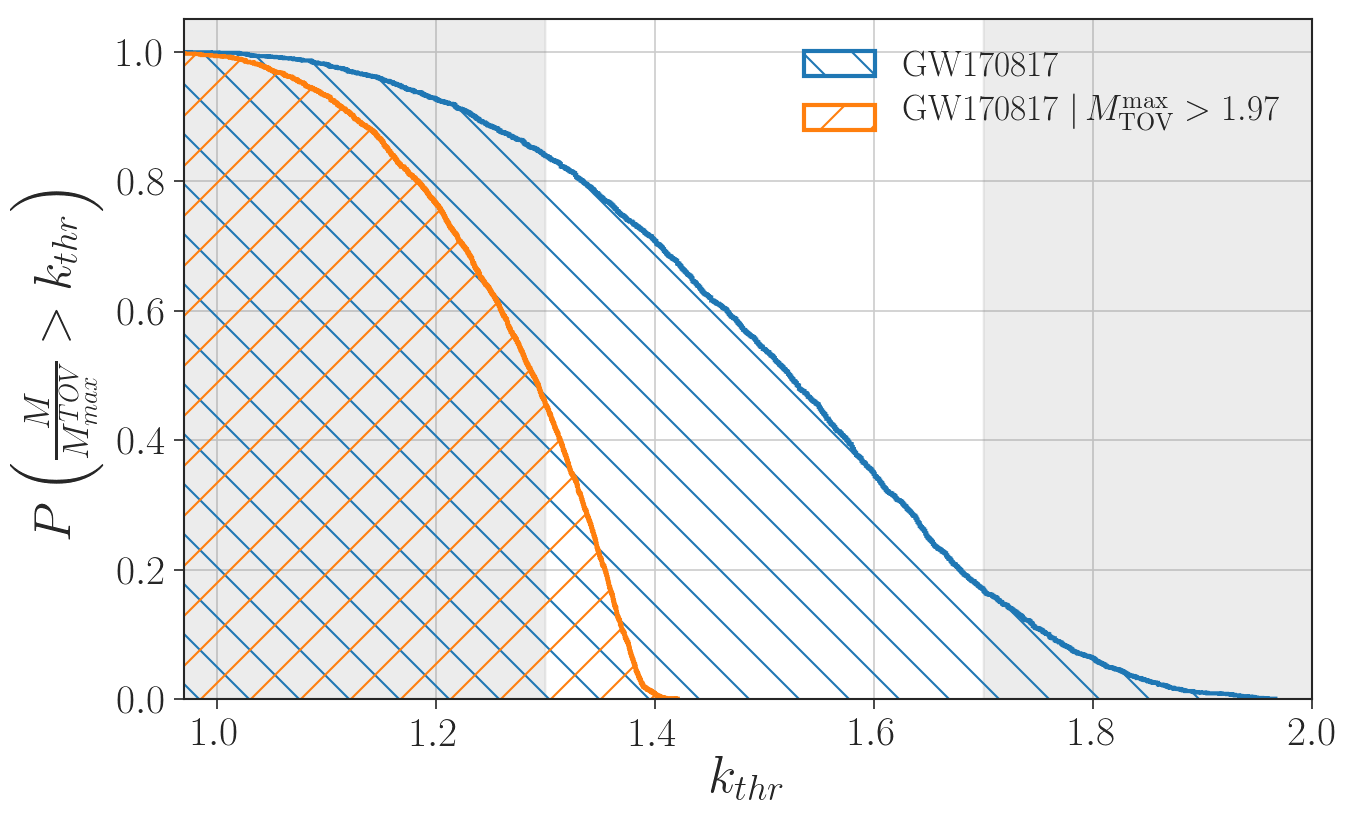}
    \caption{Prompt-collapse analysis of GW170817 based on threshold
      tidal parameter method, with and without a hard $\Mmax$ constraint at $1.97$~$\Mo$.
      Left: Joint posteriors in the $M$-$\Mmax$ plane when analysing with (orange) and without (blue) a prior cut on $\Mmax$. Dark (light) colored points lie above (below) the mass threshold of prompt collapse.
      Contours of $\kthr$ within the typical range $[1.3,1.7]$ are shown as gray shaded regions.
    Right: Probability of prompt collapse as a function of $\kthr$ with and without the $\Mmax$ constraint (before making use of Eq.~\eqref{eq:k_fit}).}
    \label{fig:mmax}
\end{figure*}

First, for the threshold-mass method we show in Figure~\ref{fig:mmax} the
joint posterior of total mass $M$ and the threshold mass $\Mthr$ (left)
as well as the cumulative distribution function of their ratio $M/\Mthr$ (right), obtained 
with the theshold mass analysis with and without the constraint
$\Mmax \ge 1.97$~$\Mo$.
The latter plot should be interpreted as the probability of prompt collapse as a
function of $\kthr$, if we pretented to be totally agnostic on $\kthr$.
Without the maximum mass constraint, the collapse probability ranges from 
$P_\text{PC} \sim0.2$ to $P_\text{PC} \sim0.85$ for the expected range
of $\kthr$ (see orange line and white region in the plot).
Including the constraint $\Mmax \ge 1.97$~$\Mo$ strongly disfavours
a prompt collapse: 
$P_\text{PC} =0$ if $\kthr> 1.4$, growing up to $P_\text{PC}
\sim0.5$ if $\kthr \sim 1.3$, if for very soft EOS and NS 
compactness $C_\text{max}\sim 0.33$.

\begin{figure}[ht]
  \centering 
  \includegraphics[width=0.49\textwidth]{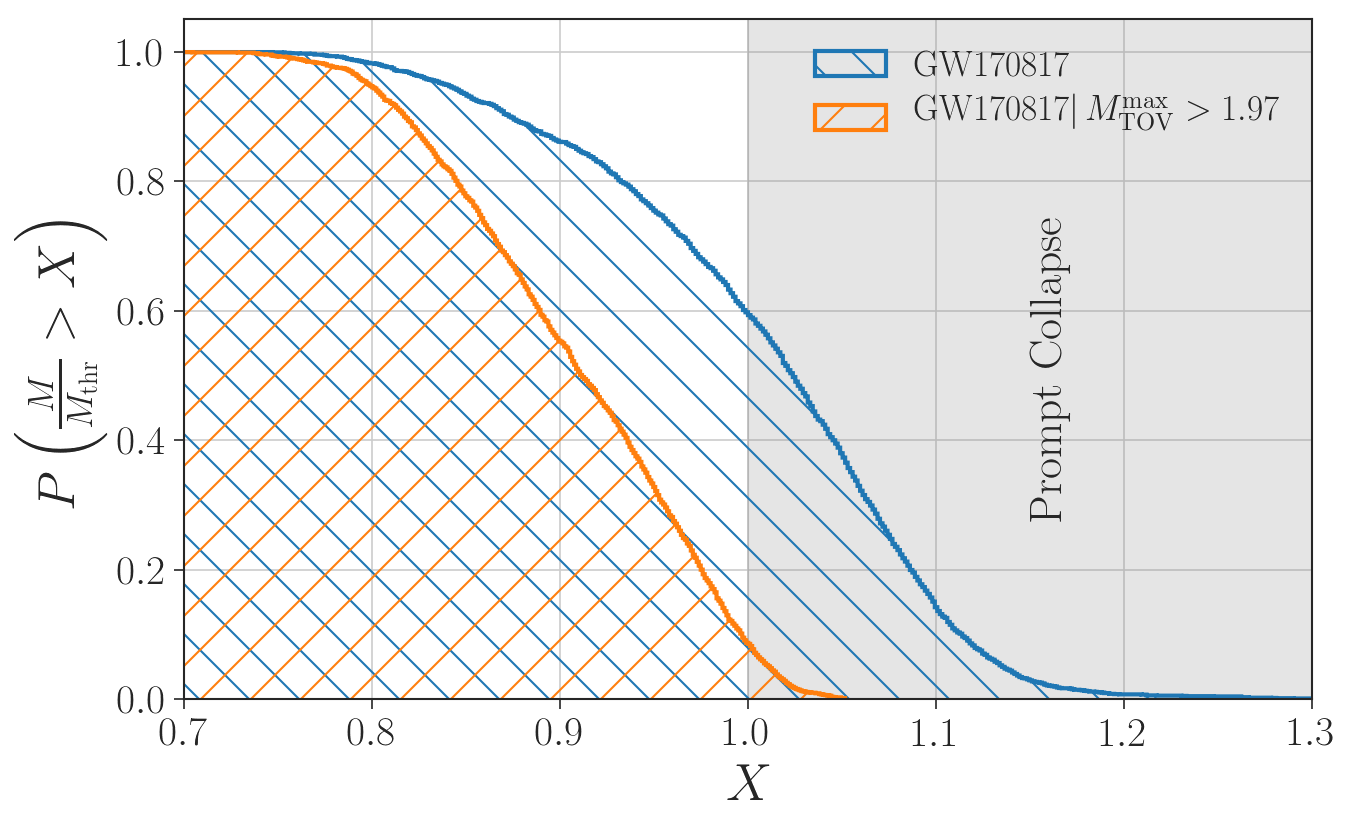}
    \caption{Cumulative posterior distribution on the ratio $M/\Mthr$. The fraction of the posterior that lies above unity gives the probability of prompt collapse with (blue) and without (orange) a constraint of $\Mmax \ge 1.97$~$\Mo$.}
    \label{fig:Ppc}
\end{figure}

\begin{figure}[t]
  \centering 
  \includegraphics[width=0.49\textwidth]{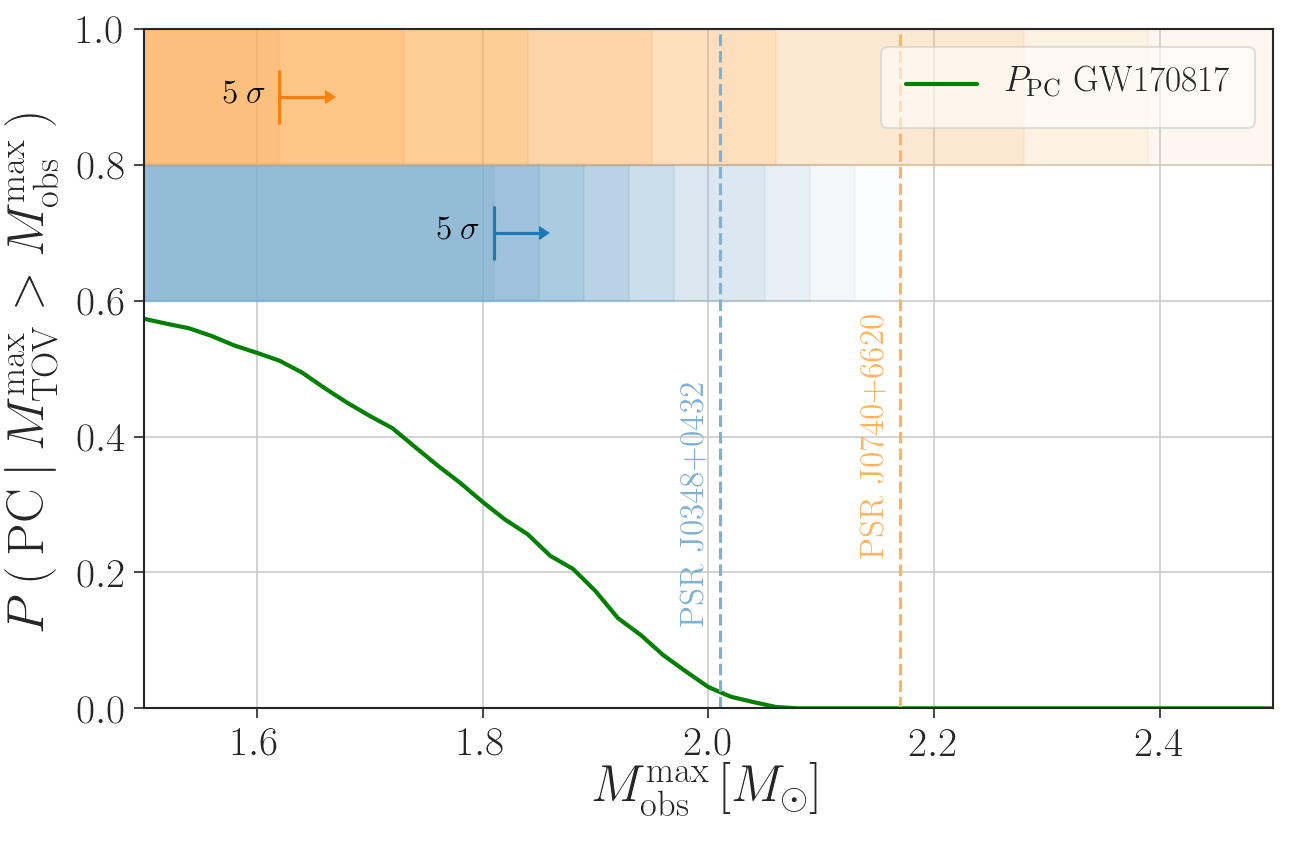}
    \caption{The probability of prompt collapse for GW170817 as a function of the heaviest observed NS mass. The Gaussian mass measurements of PSR J0348+0432 and PSR J0740+6620 are shown in the shaded regions.}
    \label{fig:Ppc_Mmax}
\end{figure}

However, $\kthr$ is not an independent unknown parameter; using the results
of Sec.~\ref{sec:Mthr_estimate} we estimate the value of $\kthr$ and $\Mthr$
from the EOS parameters $\vec{\gamma}$.
The resulting posterior of $M/\Mthr$ is plotted as a cumulative distribution
in Fig.~\ref{fig:Ppc}.
Here too, we find a significant difference between the analyses with
and without the $\Mmax$ constraint, that estimate the prompt collapse probability
at $0.09$ and $0.59$ respectively.
The reason is that the $\Mmax$ constraint removes part of the EOS parameter space that is too soft
to support a NS mass of $1.97$~$\Mo$ (and will most likely predict a prompt collapse).
The effect on $P_{\rm PC}$ is significant, since the recovered binary parameters
of GW170817 happen to lie close to the prompt-collapse threshold. 


\begin{figure}[t]
  \centering 
    \includegraphics[width=0.45\textwidth]{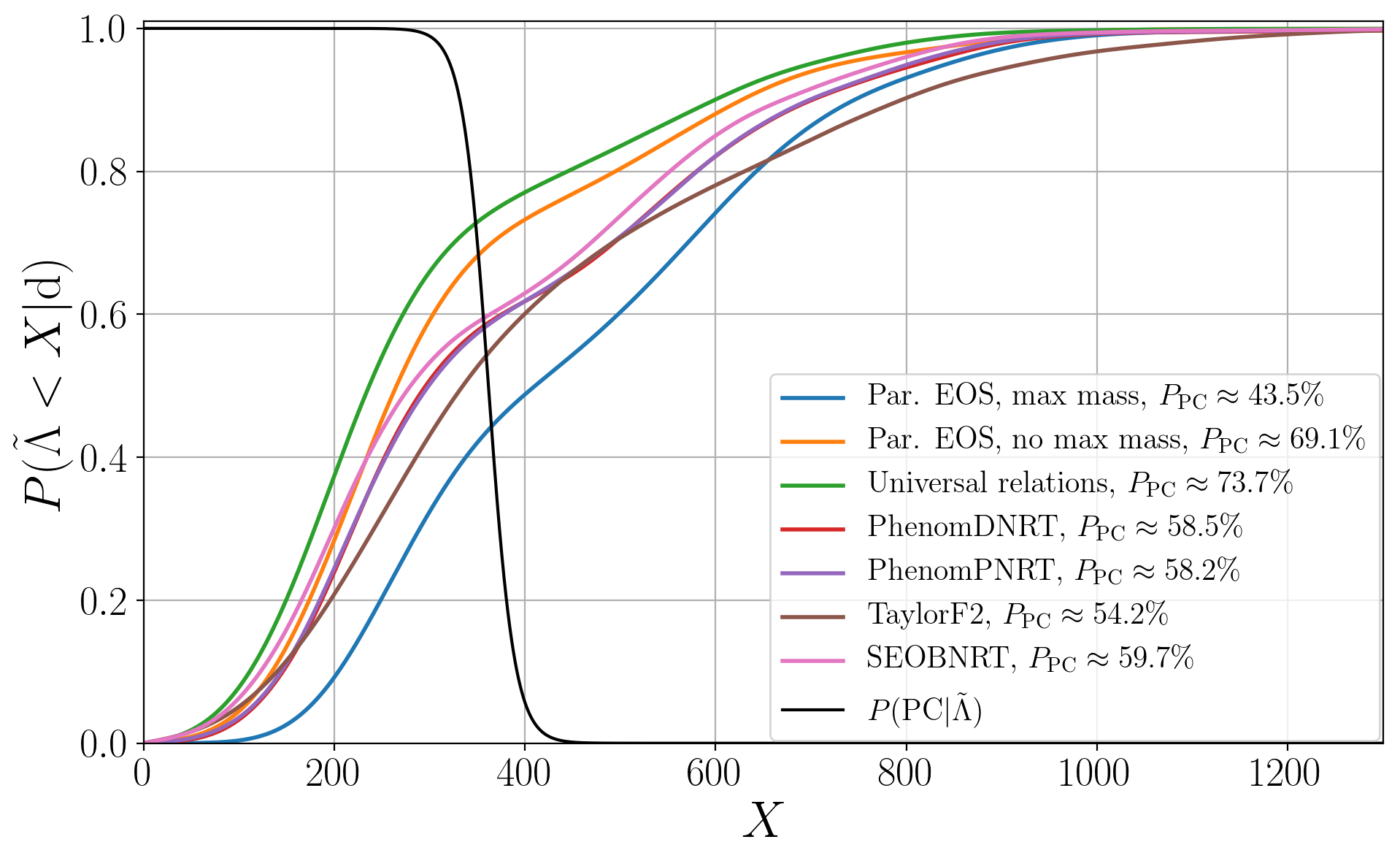}
    \caption{Probability of prompt collapse for GW170817 based on the threshold-$\tilde{\Lambda}$ method for different analysis set-ups.
      Colored curves plot the cumulative posterior probaiblity distribution for $\tilde{\Lambda}$.
      The solid black sigmoid curve gives the prior probability of prompt collapse at each value of $\tilde{\Lambda}$, based on NR simulations.
    The prompt-collapse probability can be visually estimated by the value of each curve as it crosses the transition region.}
  \label{fig:Lam_pc}
\end{figure}

Fig.~\ref{fig:Lam_pc} shows the prompt collapse probability obtained
with the threshold-$\tilde\Lambda$ method (cf. Fig.~\ref{fig:inj:lamb}, left panel)
We find a prompt collapse probability between $P_\text{PC} \sim43\%$
and $74\%$, depending on the waveform approximant used for the
analysis and on the inference method employed. The data from the EOS inference
employed also in the threshold-mass analysis give the smalleest prompt
collapse probability as a result of imposing the maximum mass
constraint. If the constraint is relaxed the probability grows to
$69\%$. All the analysis performing inference on $\tilde\Lambda$ give
prompt collapse probability between $54\%$ and $60\%$; the waveform
approximants estimating the largest $\tilde\Lambda$ clearly give the
smaller probabilities.
The largest prompt collapse probability is obtained using the EOS-insensitive
relations in the $\Lambda_{s}$ inference and employing the $\tilde{\Lambda}$ threshold.
This can be understood as the combined effect of using {\IMRP} as a template waveform
model, which tends to introduce a systematic bias favouring prompt collapse (see discussion
in Appendix~\ref{app:WfSys}) and not having a constraint on $\Mmax$.




\begin{table}[t]
  \centering    
  \caption{Summary of GW170817 results derived with the threshold-mass and the threshold-$\tilde{\Lambda}$ methods: probability of prompt collapse for different sets of analyses published by the LVC.}

  \begin{tabular}{ccccc}        
    \hline
    \hline
    Method & Inferred parameters & Approximant & Ref. & $P_\text{PC}$ \\
    \hline
    $\Mthr$ & $\vec{\gamma}$ & {\IMRP} & \cite{Abbott:2018exr} & 0.59 \\
    $\Mthr$ & $\vec{\gamma} | \Mmax \ge 1.97 \Mo$ & {\IMRP} & \cite{Abbott:2018exr} & 0.09 \\
    \hline
    $\tilde\Lambda_\text{thr}$ & $\vec{\gamma}$ & {\IMRP} & \cite{Abbott:2018exr} & 0.69 \\
    $\tilde\Lambda_\text{thr}$ & $\vec{\gamma} | \Mmax \ge 1.97 \Mo$ & {\IMRP} & \cite{Abbott:2018exr} & 0.44 \\
    $\tilde\Lambda_\text{thr}$ & $\tilde\Lambda$ & {\TF2} & \cite{Abbott:2018wiz} & 0.54 \\
    $\tilde\Lambda_\text{thr}$ & $\tilde\Lambda$ & {\IMRP} & \cite{Abbott:2018wiz} & 0.58 \\
    $\tilde\Lambda_\text{thr}$ & $\tilde\Lambda$ & {\IMRD} & \cite{Abbott:2018wiz} & 0.59 \\
    $\tilde\Lambda_\text{thr}$ & $\tilde\Lambda$ & {\SEOB} & \cite{Abbott:2018wiz} & 0.60 \\
    $\tilde\Lambda_\text{thr}$ & $\Lambda_{s}$ & {\IMRP} & \cite{Abbott:2018exr} & 0.74\\
    \hline\hline
  \end{tabular}
 \label{tab:GW170817res}
\end{table}

Resulting values for the probability of prompt collapse $P_{\rm PC}$ from the
above analyses are listed in Table~\ref{tab:GW170817res}. We observe that the
parameters of GW170817 are measured around the threshold region both for the
threshold-mass method and for the threshold-$\tilde\Lambda$ method.
Thus, overall there seems to be no definite answer as to whether the BNS
merger was followed by a prompt collapse to a BH.
However, if we focus on the analyses where the $\Mmax$ 
constraint can be imposed, to account for the mass measurement of PSR J0348+0432,
we see that the prompt-collapse hypothesis is strongly disfavoured.

We also point out that the GW170817 inference of tidal effects using various point-mass waveform approximants
combined with {\tt NRTidal} gives posteriors with a bimodal
distribution peaked around $\tilde\Lambda\sim200$ and
$\tilde\Lambda\sim600$ and support up to $\tilde\Lambda\sim800$; while
using {\TF2} and EOB approximants it gives a single broader peak at
$\tilde\Lambda\sim300$ \cite{Abbott:2018wiz,LIGOScientific:2018mvr}. 
Independent analysis confirm these findings
\cite{De:2018uhw,Dai:2018dca,Radice:2018ozg}.





\section{Conclusion}
\label{sec:conc}


We proposed two methods to infer prompt black hole
formation from the analysis of the inspiral gravitational wave of a
binary neutron merger. 
Both methods rely on numerical-relativity models of the prompt
collapse threshold for quasicircular and nonspinning binary neutron
star merger.
The methods are validated with a set of 17 injection and
recovery experiments, and verified against data from numerical relativity
simulations. All the signals were correctly recovered with the 
exception of few cases close or at the collapse threshold. Improving
such cases will require more precise numerical relativity models and
simulations. 
We conclude that our analysis could be robustly applied to
GW170817-like signals (single events) captured by advanced LIGO-Virgo at designed
sensitivity.
We also point out that waveform systematics may introduce
important biases in the near-threshold region.

Application of these two methods on the GW170817 data gives no
definitive answer to whether the BNS merger was followed by a prompt
collapse into a BH, as the recovered masses and tidal parameters of
the binary appear to lie in the vicinity of the threshold.
However, if a constraint is applied on the maximum irrotational NS
mass supported by the EOS, that is compatible with the mass
measurements of PSR J0348+0432 and PSR J0740+6620, then we observe a
strong preference against the prompt collapse hypothesis. 



\begin{acknowledgments}
  The authors thank the LIGO-Virgo matter group for discussions, 
  Liang Dai and Kenta Hotokezaka for sharing some data and information.  
  MB, SB, FZ acknowledge support by the EU H2020 under ERC Starting
  Grant, no.~BinGraSp-714626.  
  DR acknowledges support from a Frank and Peggy Taplin Membership at the
  Institute for Advanced Study and the
  Max-Planck/Princeton Center (MPPC) for Plasma Physics (NSF PHY-1804048).
  Data analysis for this paper was performed on the supercomputer ARCCA;
  we are grateful for computational resources provided by Cardiff University,
  and funded by STFC grant ST/I006285/1.
  Data analysis was performed on the Virgo ``Tullio'' server 
  at Torino supported by INFN.
  Numerical relativity simulations were performed 
  on the supercomputer SuperMUC at the LRZ Munich (Gauss project
  pn56zo), 
  on supercomputer Marconi at CINECA (ISCRA-B project number HP10BMHFQQ); on the supercomputers Bridges, Comet, and Stampede 
  (NSF XSEDE allocation TG-PHY160025); on NSF/NCSA Blue Waters (NSF
  AWD-1811236; on ARA cluster at Jena FSU.
\end{acknowledgments}

\bibliography{references}

\appendix



\section{Prompt collapse threshold from NR data}
\label{app:pckthr}

We collect NR data for the prompt collapse threshold $\kthr$
from
\cite{Hotokezaka:2011dh,Bauswein:2013jpa,Zappa:2017xba,Koppel:2019pys} into Tab.~\ref{tab:kthr:data}. 
\begin{table}[h]
  \centering    
  \caption{Numerical relativity data used for determinig the prompt
    collapse mass threshold coefficient $\kthr$.}
  \begin{tabular}{ccccccccc}        
    \hline\hline
    EOS & $C_\text{max}$ & $\kthr$ & $\delta \kthr$ &$\Mmax$& $\Mthr$& Ref\\
     &  & &&$[\Mo]$& $[\Mo]$& \\    
\hline
APR4    &0.329	&1.243	&0.023 &2.21&2.80&\cite{Hotokezaka:2011dh}\\
SLy	&0.307	&1.342	&0.024 &2.06&2.76&\cite{Hotokezaka:2011dh}\\	 
H3      &0.224	&1.566	&0.056 &1.79&2.90&\cite{Hotokezaka:2011dh}\\
H4	&0.258	&1.452	&0.025 &2.03&2.95&\cite{Hotokezaka:2011dh}\\
ALF2	&0.260	&1.414	&0.024 &1.99&2.81&\cite{Hotokezaka:2011dh}\\
\hline
NL3	&0.307	&1.380	&0.018 &2.79&3.85&\cite{Bauswein:2013jpa}\\
GS1	&0.306	&1.400	&0.018 &2.75&3.85&\cite{Bauswein:2013jpa}\\
LS375	&0.325	&1.347	&0.018 &2.71&3.65&\cite{Bauswein:2013jpa}\\
DD2	&0.300	&1.384	&0.021 &2.42&3.35&\cite{Bauswein:2013jpa}\\
Shen	&0.250	&1.554	&0.023 &2.22&3.45&\cite{Bauswein:2013jpa}\\ 
TM1	&0.260	&1.561	&0.023 &2.21&3.45&\cite{Bauswein:2013jpa}\\
SFHx	&0.292	&1.432	&0.023 &2.13&3.05&\cite{Bauswein:2013jpa}\\
GS2	&0.262	&1.555	&0.024 &2.09&3.25&\cite{Bauswein:2013jpa}\\
SFHo	&0.294	&1.432	&0.024 &2.06&2.95&\cite{Bauswein:2013jpa}\\
LS220	&0.284	&1.495	&0.025 &2.04&3.05&\cite{Bauswein:2013jpa}\\
TMA	&0.247	&1.609	&0.025 &2.02&3.25&\cite{Bauswein:2013jpa}\\
IUF	&0.255	&1.564	&0.026 &1.95&3.05&\cite{Bauswein:2013jpa}\\
\hline
LS220	&0.284	&1.445	&0.024 &2.04&2.95&\cite{Zappa:2017xba}\\	
BHB$\Lambda\phi$ 	&0.268	&1.469	&0.047 &2.11&3.10&\cite{Zappa:2017xba}\\
ALF2	&0.260	&1.444	&0.063 &1.99&2.86&\cite{Zappa:2017xba}\\
H4	&0.258	&1.529	&0.049 &2.03&3.10&\cite{Zappa:2017xba}\\
SLy	&0.307	&1.395	&0.061&2.06&2.86&\cite{Zappa:2017xba}\\
\hline
BHB$\Lambda\phi$     &0.268	&1.503	&0.024 &2.11&3.17&\cite{Koppel:2019pys}\\
DD2	&0.300	&1.364	&0.021&2.42& 3.30&\cite{Koppel:2019pys}\\
SFHo	&0.294	&1.391	&0.024 &2.06&2.87&\cite{Koppel:2019pys}\\
TM1	&0.260	&1.520	&0.023 &2.21&3.36&\cite{Koppel:2019pys}\\
    \hline\hline
  \end{tabular}
 \label{tab:kthr:data}
\end{table}

In the first three references, the collapse threshold is estimated by
performing simulations with a given EOS and different masses and then
linearly interpolating between the two simulations that braket the
threshold.
The uncertainty on the $\kthr$ is thus determined
not only by the grid resolution of the simulations but also by how
close the two simulation bracket the threshold.
Typical relative errors
are at the level of 5\% although, surprisingly, detailed grid resolution
studies for this application are missing. 
In \cite{Koppel:2019pys} the collapse threshold for a specific EOS is
determined as the binary mass for which the merger remnant collapses
over the free-fall timescale of a $\Mmax$ NS. The
actual value of $\Mthr$ is computed using an extrapolation of an exponential fit of
the results obtained by a few prompt collapse simulations. The
reported error is also the one obtained by the exponential
fit. Despite providing consistent results, the error estimate is
qualitatively different from the other approaches and
counterintuitively the fewer the simulations the smaller the
uncertainties. We thus select the $\Mthr$ for which at least
three simulations were performed and we assign to all of them the
relative error obtained by their DD2 model, since this is the model with
more points. We stress that this relative error is comparable to the
smallest relative errors used in the other works. 

\begin{figure}[h]
  \centering 
  \includegraphics[width=0.49\textwidth]{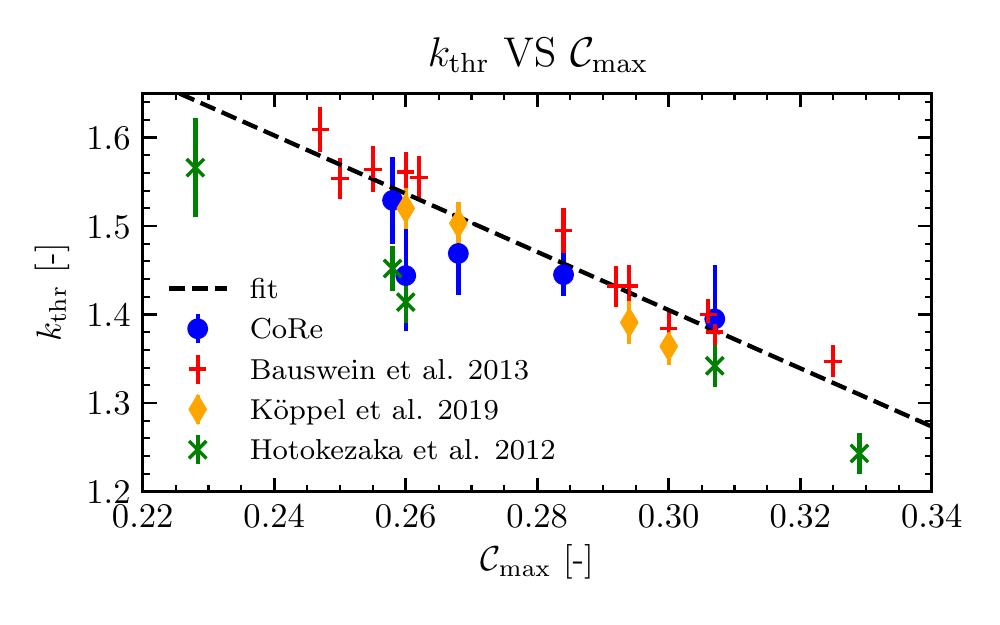}
    \includegraphics[width=0.49\textwidth]{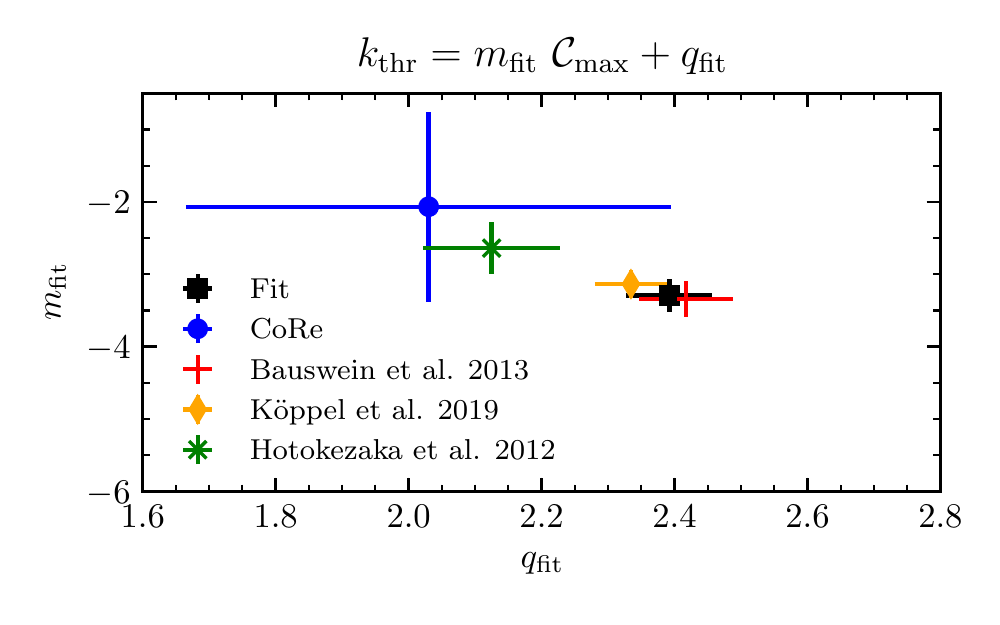}
    \caption{Numerical relativity fits of the prompt collapse threshold $\kthr$. Top: $k_{\rm thr}$
    as a function of the maximum compactness $C_{\rm max}$, see Table~\ref{tab:kthr:data}. The black line represents the fit of the results reported in \cite{Zappa:2017xba,Koppel:2019pys,Bauswein:2013jpa}. Bottom: coefficients of the linear fits (with errors) obtained by the single datasets and by the dataset used to produce the fit used in this work.}
    \label{fig:ktrhfit}
    %
\end{figure}

The data are plotted as a function of the maximum compactness and
shown in the top panel of Fig.~\ref{fig:ktrhfit}. All the data show an approximate
linear correlation with $C_\text{max}$ although part of the data of
\cite{Hotokezaka:2011dh} appear to systematically underestimate the
threshold with respect to the other datasets. 
We perform several linear fits combining the datasets and propagating the uncertainties. The
fit coefficients and their errors are reported in the bottom
panel of Fig.~\ref{fig:ktrhfit}. The different sets of coefficients
are essentially compatible with each other, and the errors become
smaller when more points are included.
The equation we use in the main body (Eq.~\eqref{eq:k_fit}) is the best fit given by
the combination of the data of \cite{Bauswein:2013jpa,Zappa:2017xba,Koppel:2019pys}. 



\section{Waveform systematics}
\label{app:WfSys}

\begin{figure*}[h!]
  \centering 
    \includegraphics[width=0.49\textwidth]{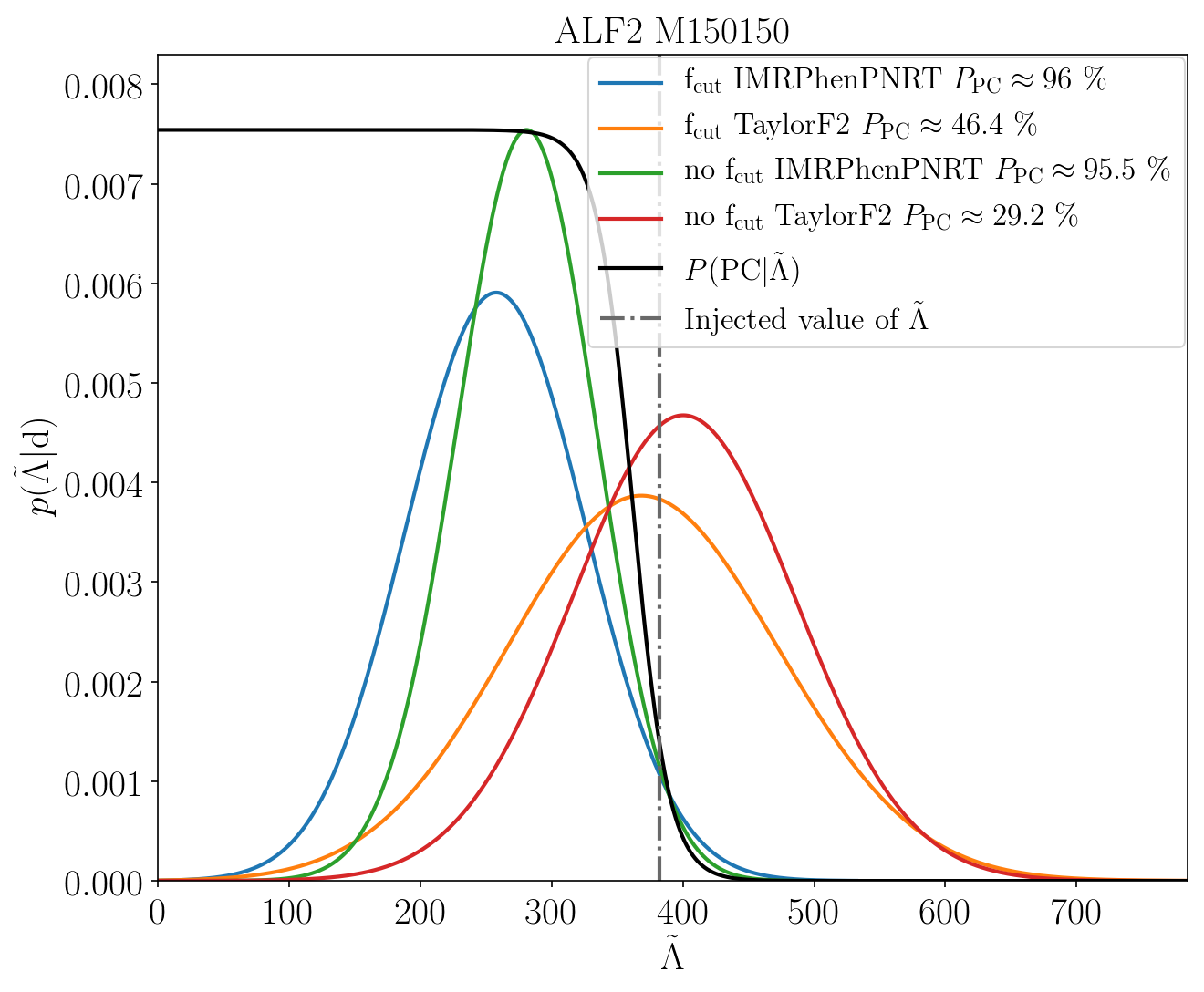}
    \includegraphics[width=0.49\textwidth]{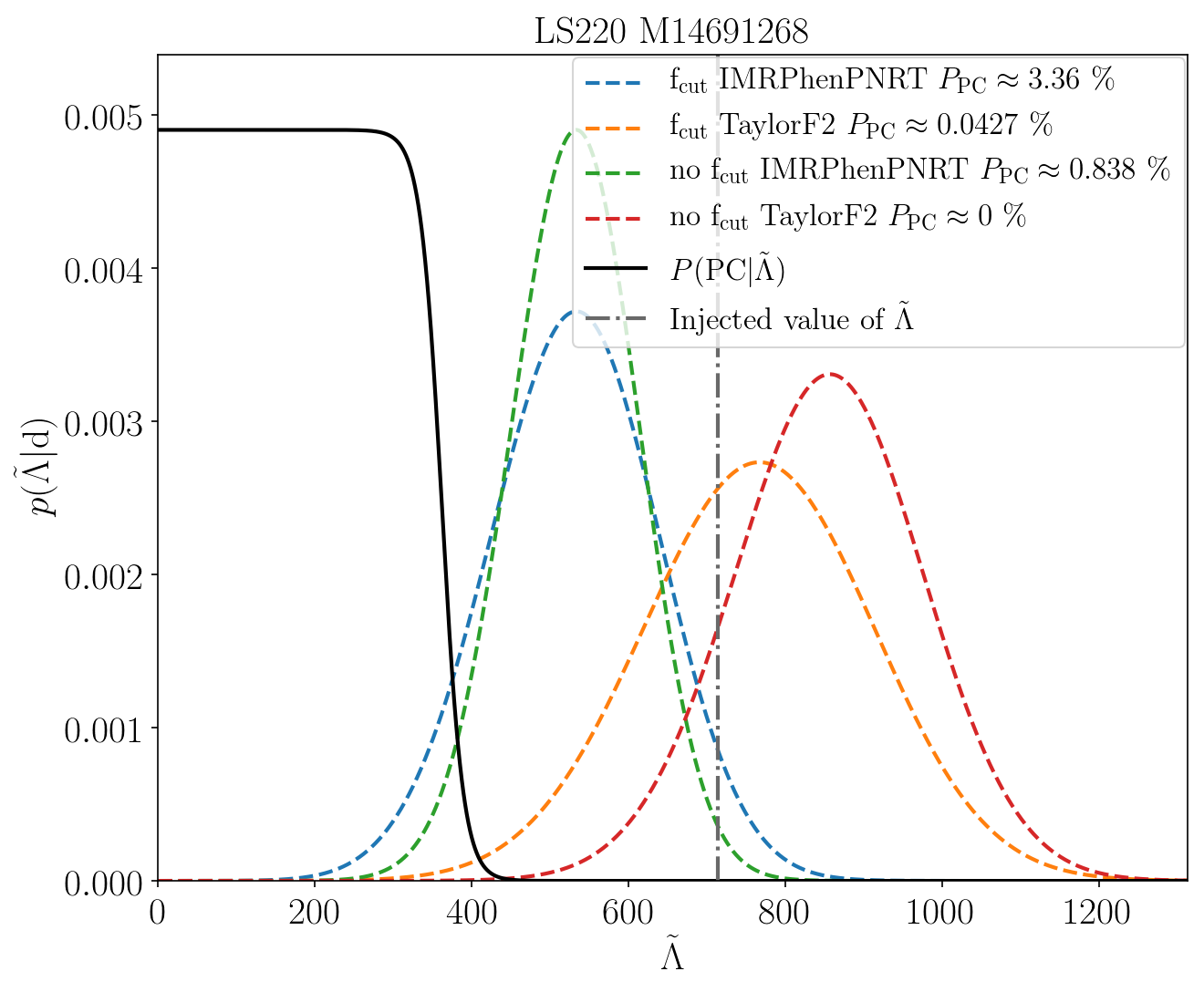}
    \caption{Recovery of {\TEOB}  $\tilde\Lambda$ with {\TF2} and {\IMRP} and two
      maximal cut-off frequencies $1024$~Hz and $2048$~Hz for
      representative injections.
      In our experiments {\IMRP} systematically underestimates the injected
      {\TEOB} $\tilde\Lambda$, while {\TF2} with cut-off
      frequency $1024$~Hz give the most consistent results.}
    \label{fig:TF2_Pheno_bias:Lambdathr1}
\end{figure*}

\begin{figure*}[h!]
  \centering 
    \includegraphics[width=0.49\textwidth]{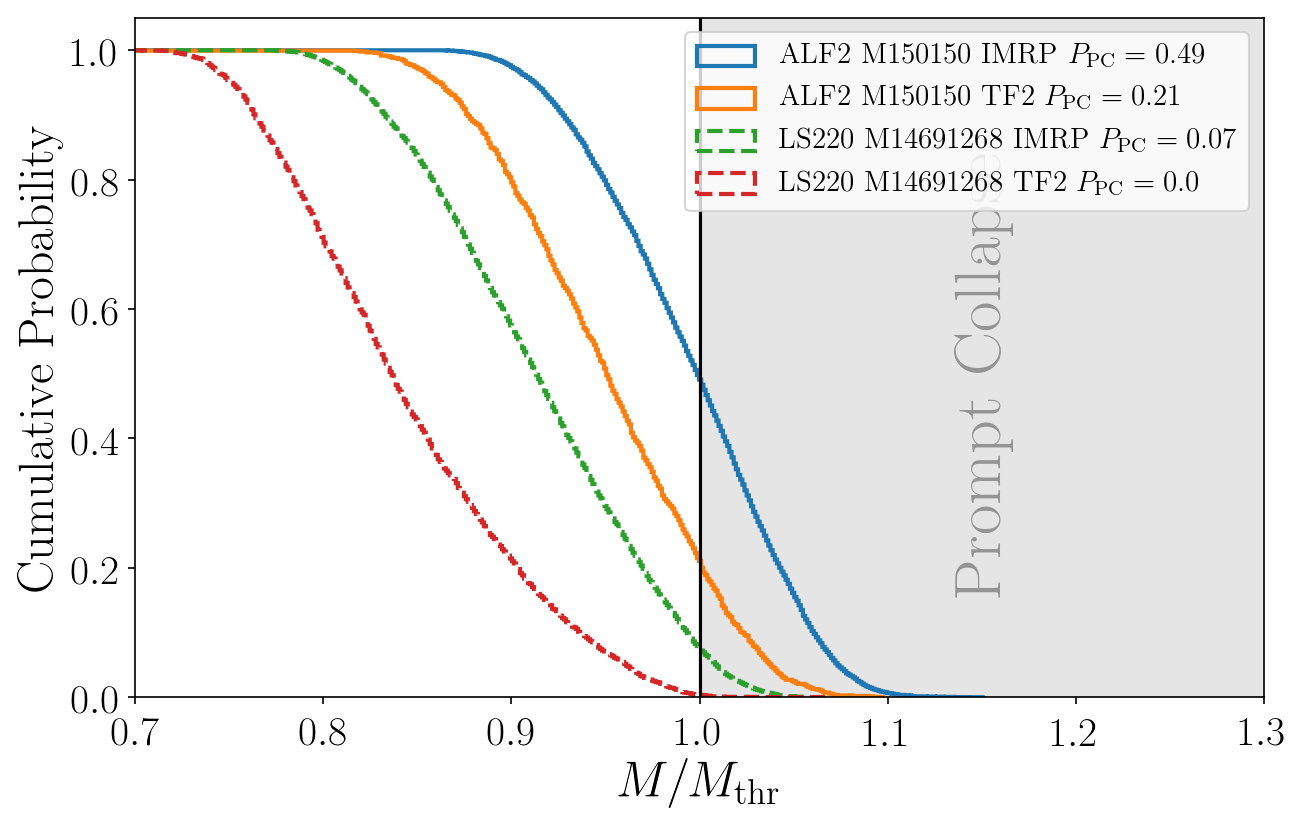}
    \includegraphics[width=0.49\textwidth]{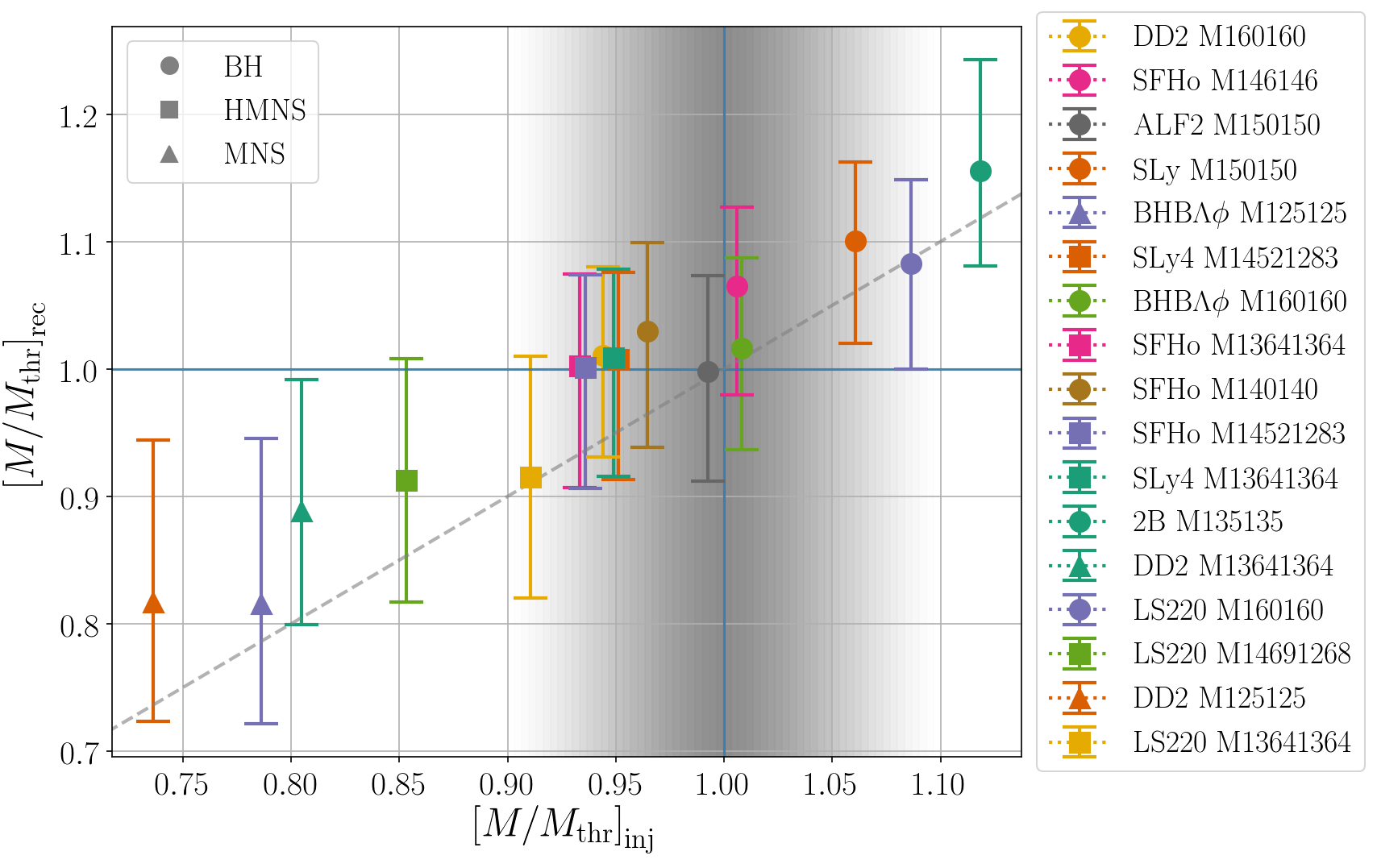}
    \caption{Left: Recovery of {\TEOB} mass relative to threshold mass with {\TF2} and {\IMRP} for the two runs shown in Fig.~\ref{fig:TF2_Pheno_bias:Lambdathr1}. Waveform systematics may induce significant effects on the threshold mass parameter analysis. Right: Summary of threshold-mass analysis on the simulated signals of Table~\ref{tab:NRSim} using {\IMRP}.
    }
    \label{fig:TF2_Pheno_bias:Massthr}
\end{figure*}

Our injection experiments highlight systematics biases in the
recovery of the {\TEOB} waveforms when using the {\IMRP} as our template model, while 
results are overall consistent when {\TF2} is employed with or without
a high cut-off frequency of $1024$~Hz.
Representive measurements of the tidal parameter $\tilde{\Lambda}$ are shown
in Fig.~\ref{fig:TF2_Pheno_bias:Lambdathr1} for two injections with different EOS.
A similar bias between {\TEOB} and {\IMRP} is visible in the injections of
\cite{Dudi:2018jzn,Messina:2019uby} performed at SNR 100, 
but so far this has not been systematically investigated nor explained.
We plan to do so in a forthcoming publication.
Here, we use the {\IMRP} results to discuss
how waveform systematics can affect the prompt collapse inference with 
our two methods.
In the comparison plots of Fig.~\ref{fig:TF2_Pheno_bias:Lambdathr1} we show how
analyzing the same signal with different waveform models can affect the estimated
prompt collapse probability; in particular {\IMRP} tends to underestimate tidal deformabilities
and therefore overestimate $P_{\rm PC}$. A similar effect is observed in the threshold-mass
results for the same injections in the left panel of Fig.~\ref{fig:TF2_Pheno_bias:Massthr}, where
$M/\Mthr$ is overestimated by {\IMRP} (and thus so is $P_{\rm PC}$).
Overall, we find the significance of waveform systematics to be 
limited to the cases close to the collapse threshold.

\begin{figure*}[t]
  \centering 
    \includegraphics[width=0.49\textwidth]{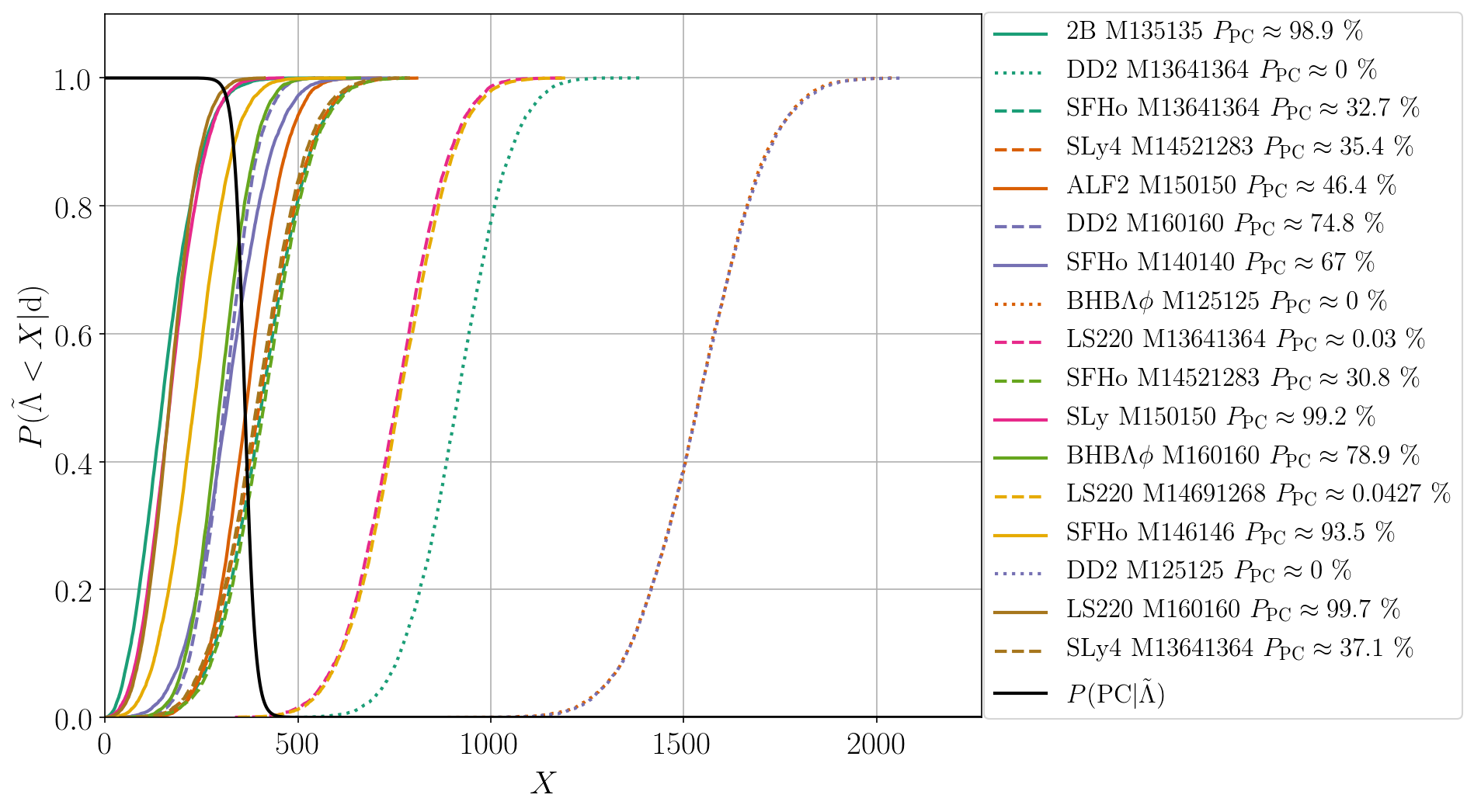}
    \includegraphics[width=0.49\textwidth]{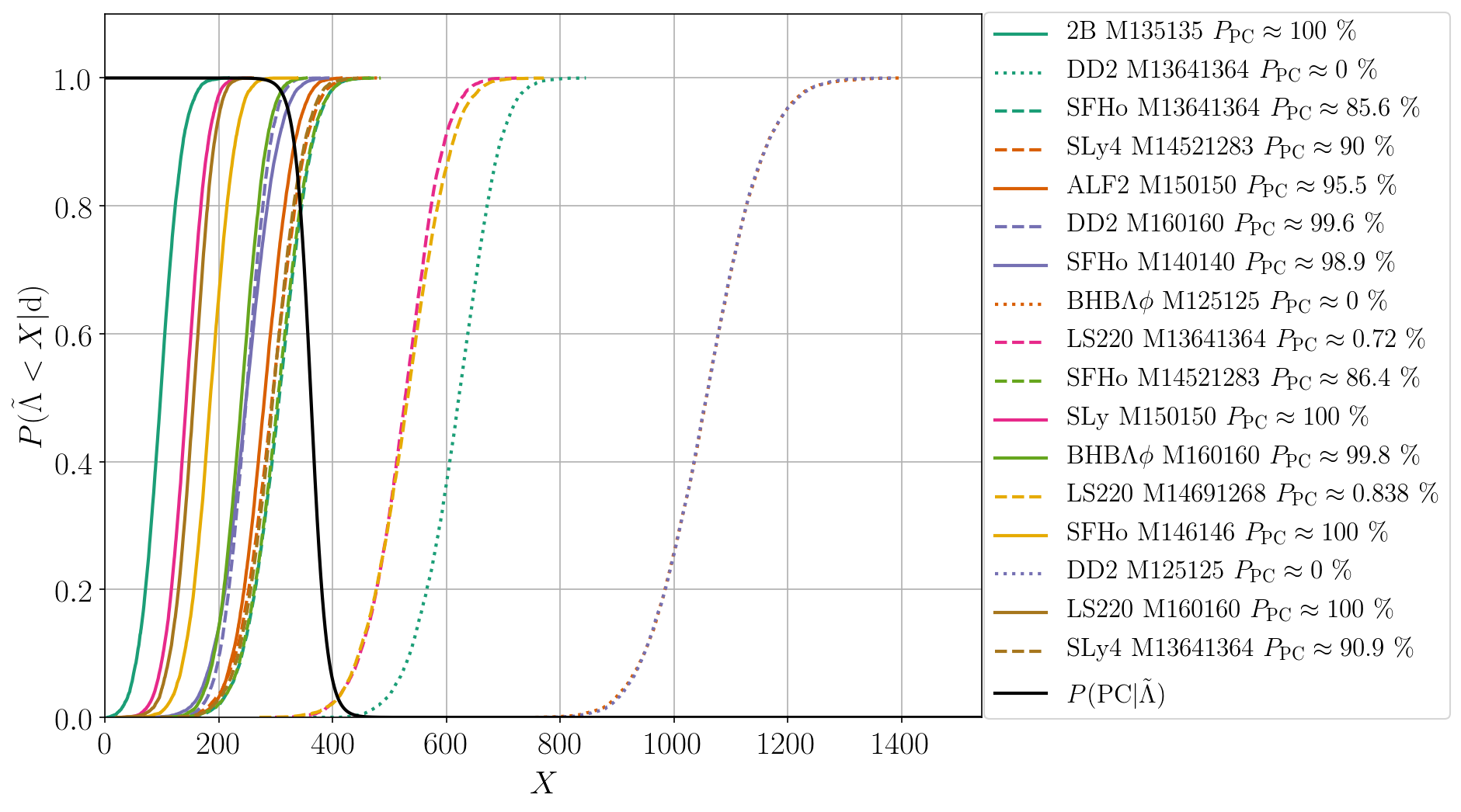}
    \caption{Waveform systematics effects on the threshold tidal
      parameter analysis.
      Recovering with {\TF2} and cut-off $1024$~Hz (left) gives
      consistent results with the injection except for binaries with 
      $\Lambda\sim\tilde\Lambda_\text{thr}$ for which a 50-50 chance
      of prompt is returned.
      Recovering with {\IMRP} and cut-off $2048$~Hz (right) gives
      consistent results with the injection except for binaries with 
      $\Lambda\sim\tilde\Lambda_\text{thr}$ for it incorrectly
      favours prompt collapse.}
    \label{fig:TF2_Pheno_bias:Lambdathr2}
\end{figure*}

Figure~\ref{fig:TF2_Pheno_bias:Lambdathr2} summarizes waveform systematics
effects on the threshold tidal parameter analysis.
We find that the prompt collapse inference with {\TF2} and cut-off $1024$~Hz gives
consistent results with the injection except for binaries with 
$\Lambda\sim\tilde\Lambda_\text{thr}$.
For the DD2 1.59+1.59 BNS ($\tilde\Lambda=332$) and the two SLy binaries
($\tilde\Lambda\sim401$) the method estimates respectively a 75\% and
$\sim40$\% probability of prompt collapse while the merger result in a
HMNS. In the former case the binary is at the collapse threshold and
the HMNS is very short lived ($3$~ms). Hence it could be simply a
results of out uncertainties. In the latter case the binaries are slightly above
the collapse threshold and the prediction appears to have a genuine systematic error of the method.
Similarly, for SFHo 1.40+1.40 ($\tilde\Lambda=334$) and ALF2 1.50+1.50 ($\tilde\Lambda=382$) the method predicts
33\% and 54\% probability of producing a NS remnant while the simulations
indicate prompt BH formation.

Recovering with {\IMRP} systematically underestimates the injected
{\TEOB} $\tilde\Lambda$; the effect being worst for cut-off frequency
$1024$~Hz and minimzed by cut-off $2048$~Hz. The result can be in part
understood from the fact that the low frequency limit of the {\tt
  NRtidal} is accurate only to the leading-order post-Newtonian tidal
term \cite{Dietrich:2017aum,Dietrich:2018uni}.
The same systematic trend can be seen in the threshold-mass analysis
summarized in the right panel of Fig.~\ref{fig:TF2_Pheno_bias:Massthr},
which is more pronounced in the less compact binaries.
The errors in the prompt collapse analysis due to
the numerical fits on $\kthr$ discussed above, are now combined with
those from the waveform systematics. As a result, the
method predicts correctly the prompt collapse of ALF2 1.59+1.59 and SFHo 1.40+1.40 (thanks
to a ``cancellation'' of systematic errors) but incorrectly favours
prompt collapse for the SLy binaries.
      

\section{Effect of $\Mmax$ constraint}
\label{app:maxM}

In the threshold-mass method, sampling the EOS parameter space directly
allowed us to impose conditions on the maximum stable nonrotating NS mass,
$\Mmax$.
In this section we examine the effect that different choices of this
constraint may have on estimating the probability of prompt collapse.

\begin{figure}[h!]
  \centering 
    \includegraphics[width=0.49\textwidth]{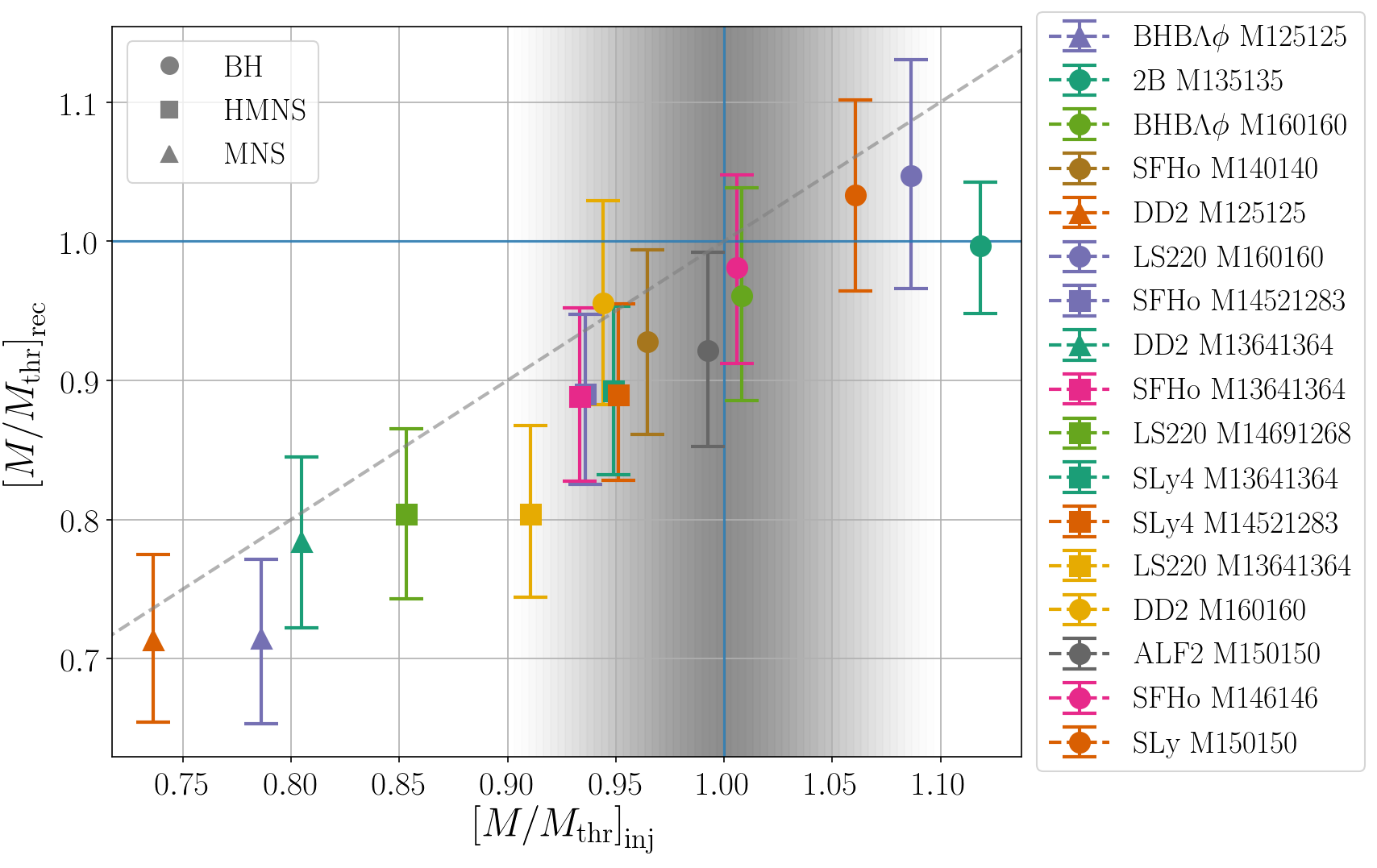}
    \caption{Effect of $\Mmax$ constraint on the threshold mass 
      parameter analysis (see Fig.~\ref{fig:inj:mmax}) using the PSR J0348+0432 mass measurement.}
    \label{fig:Mmax_summary}
\end{figure}

First we review the results of the injection study of Sec.~\ref{sec:validation}
when imposing a $\Mmax$ constraint based on the mass measurement of PSR J0348+0432.
Results are summarized in Fig.~\ref{fig:Mmax_summary}.
When comparing against Fig.~\ref{fig:inj:mmax}, we observe a systematic trend
to lower values of recovered $M/\Mthr$. This can be interpreted as a push
towards higher values of $\Mthr$, which is expected, since a soft part of the space
of EOS is effectively removed from our prior.
Note the peculiar behavior of the 2B BNS as 
a consequence of the fact that the maximum mass for that EOS
violates the prior imposed in the analysis.

\begin{figure}[h!]
  \centering 
    \includegraphics[width=0.49\textwidth]{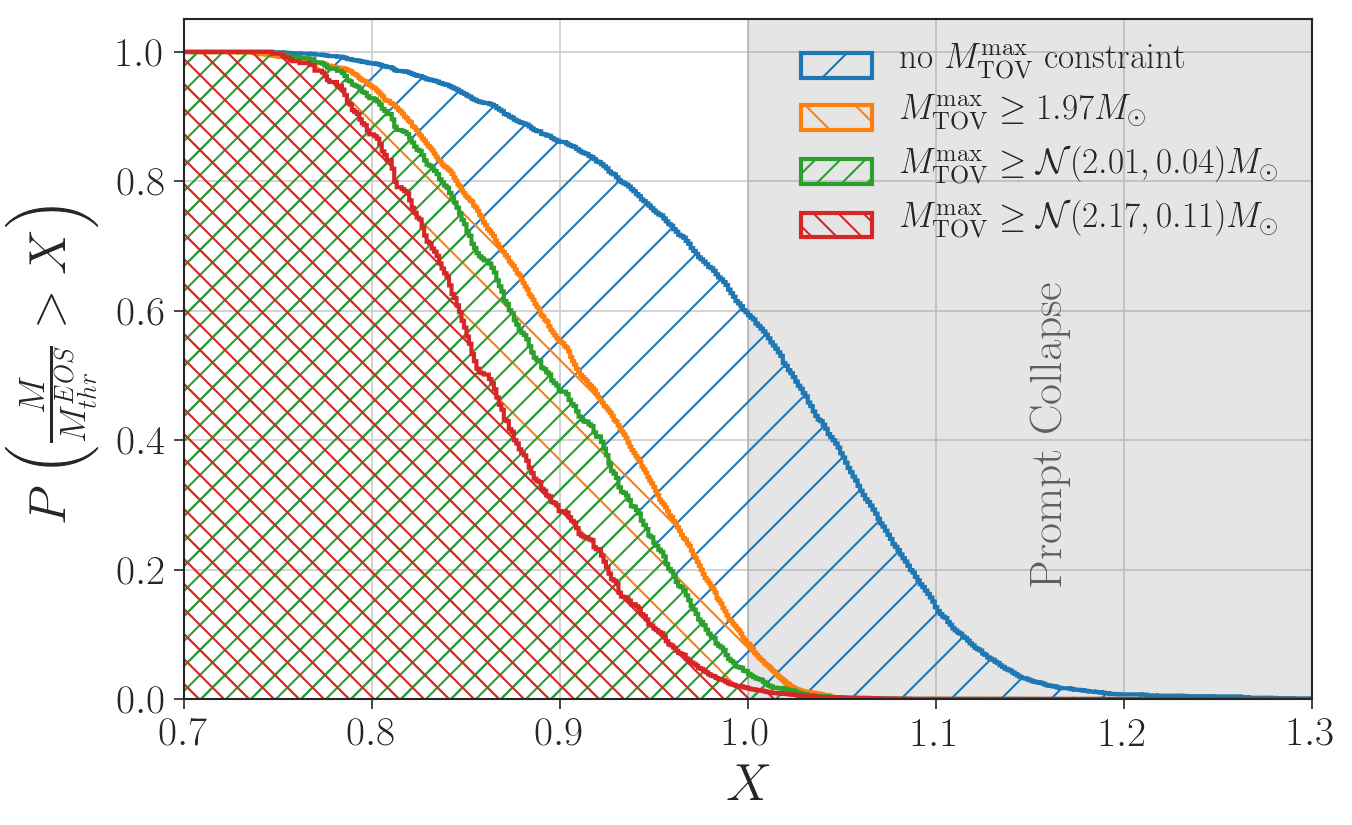}
    \caption{Cumulative distribution of the total mass $M$ divided by the threshold mass $\Mthr$ for different choices of the $\Mmax$ constraint. The value at $X=1$ gives the probability of prompt collapse.}
    \label{fig:Mmax_comp}
\end{figure}

We now move on to the analysis of GW170817 data using the spectral EOS parametrization
and consider the following choices:
\begin{itemize}
\item No constraint on $\Mmax$;
\item A hard constraint of $\Mmax \ge 1.97 \Mo$, corresponding to a
  conservative 1-$\sigma$ bound on the mass of PSR J0348+0432;
\item A probabilistic constraint based on the mass measurement of
  PSR J0348+0432, which follows the Gaussian PDF $\mathcal{N}(2.01,0.04)$;
\item A probabilistic constraint based on the recent observation of
  PSR J0740+6620, which follows the Gaussian PDF $\mathcal{N}(2.17,0.11)$;
\end{itemize}
The results are illustrated in Fig.~\ref{fig:Mmax_comp}. We find that
if the heavy-NS measurements are taken into account, the prompt-collapse probability
tends to zero (even more so than in the case of a hard cut at $1.97 \:\Mo$).

\section{EOS reconstruction}
\label{app:EOSrec}

\begin{figure*}
  \centering 
    \includegraphics[width=0.49\textwidth]{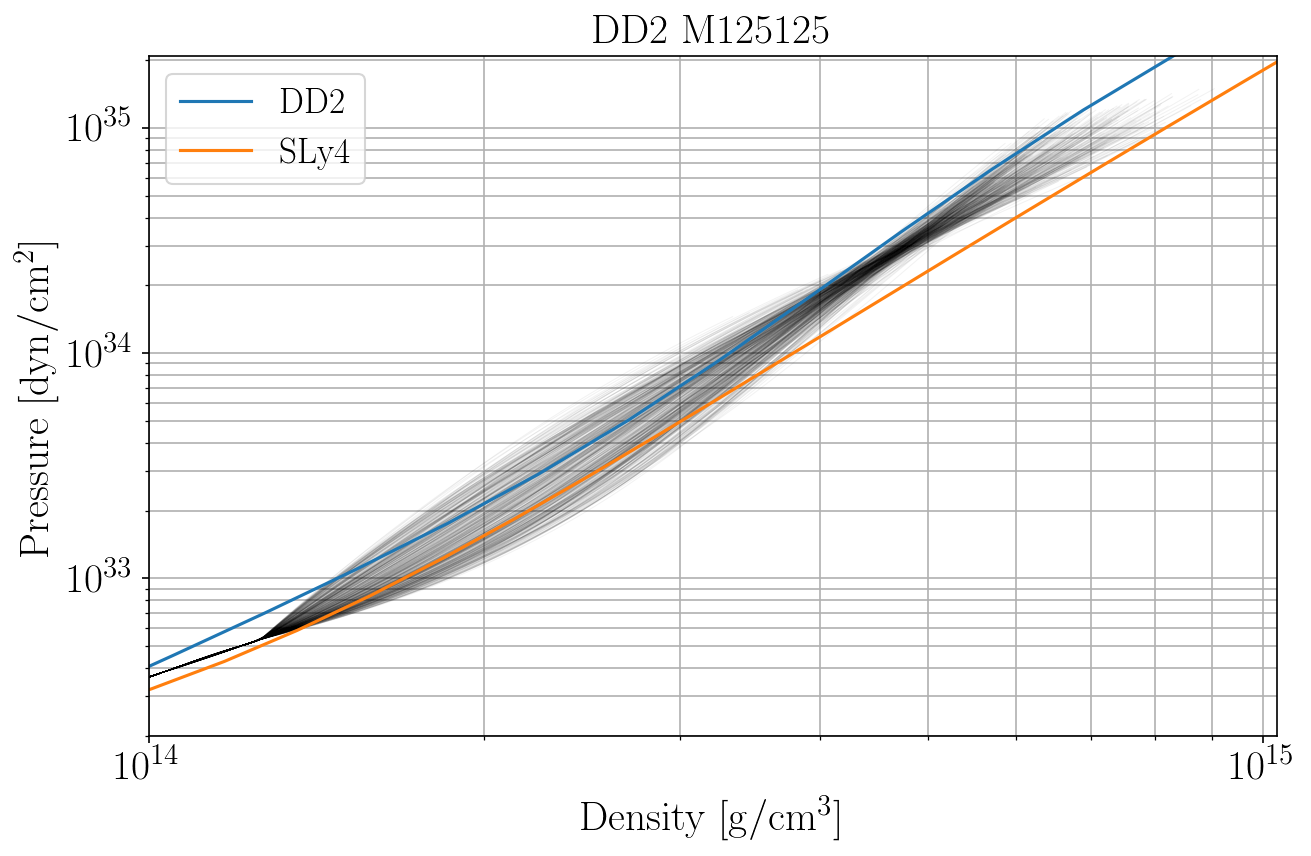}
    \includegraphics[width=0.49\textwidth]{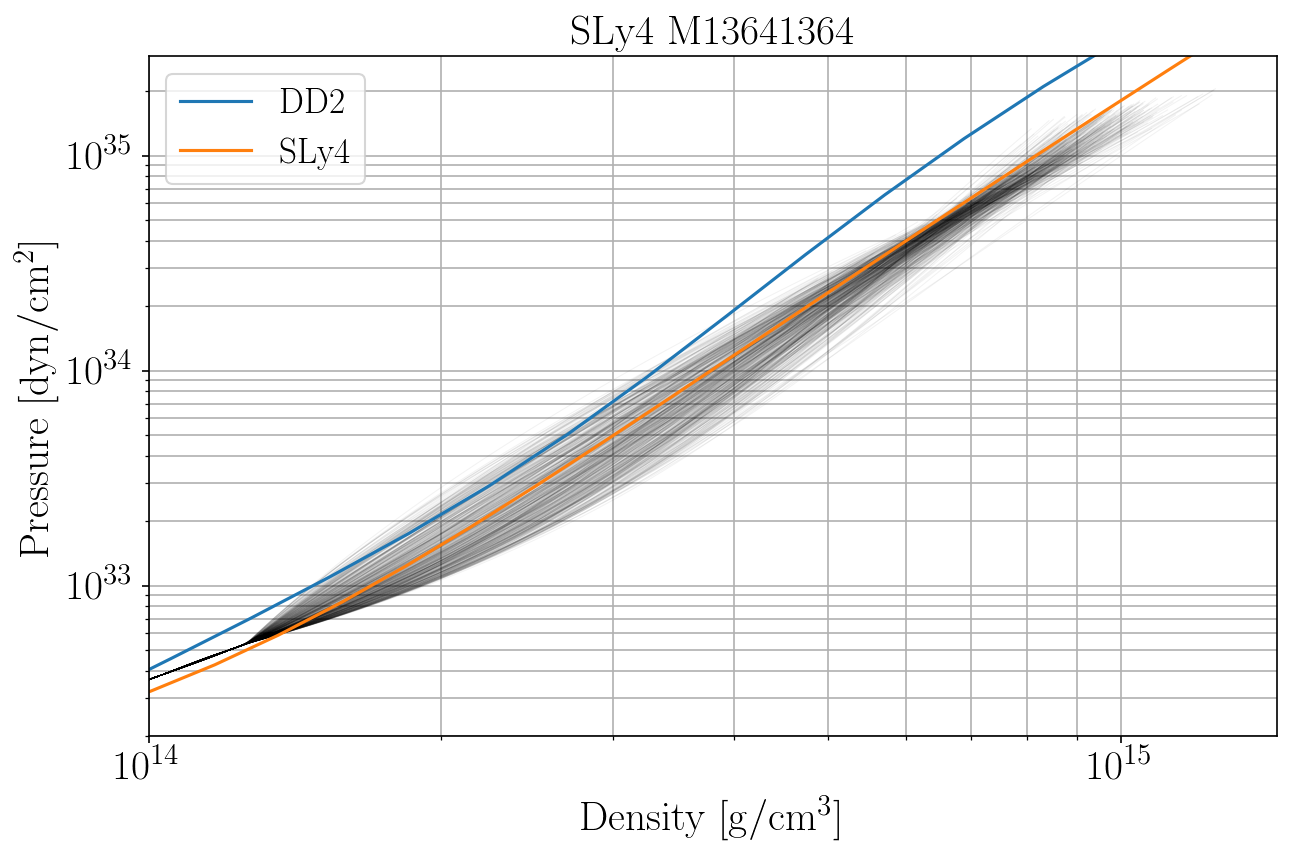}
    \caption{Reconstruction of $P(\rho)$ for the underlying EOS models DD2 (left) and SLy4 (right), from the posterior PDF on the spectral parameters $\vec\gamma$. The posteriors in the two cases converge towards the correct EOS curve, with a ``focal point'' around $2\:\rho_{\rm nuc}$. The curves end at the central density of their corresponding heaviest stable NS.}
    \label{fig:EOSrecon}
\end{figure*}

In our threshold-mass method we have employed the spectral family of~\cite{Lindblom:2010bb}
to parametrize the EOS. It is instructive to examine whether the posterior
PDF on the EOS parameters $\vec\gamma$ faithfully reconstructs the injected model,
within the margins of our measurement error. Two typical cases are illustrated in
Fig.~\ref{fig:EOSrecon}, where the reconstructed $P(\rho)$ curves are clearly distinguishable
from each other and faithfully follow the corresponding underlying model.
In particular we observe the separation becoming more clear around a ``focal point'' at
$\sim 2\:\rho_{\rm nuc}$, which happens to be close to the typical central density of the NS,
which largely determines the bulk properties of the star.

\end{document}